\documentclass[journal]{IEEEtran}
\usepackage{setspace}
\usepackage{amsfonts}
\usepackage{bm}
\usepackage{amsmath}
\usepackage{amssymb}
\usepackage{enumerate}
\usepackage{color,xcolor}
\usepackage{graphicx}
\usepackage{multirow,tabularx}
\usepackage{subfigure}
\usepackage{hyperref}
\usepackage[compress]{cite}
\usepackage{algorithm}
\usepackage{algpseudocode}
\usepackage{subfigure}
\usepackage{graphicx}
\usepackage{caption}
\usepackage{mathtools}
\newtheorem{remark}{Remark}
\usepackage{siunitx}
\ifCLASSINFOpdf
\else
\fi
\begin{document}

\title{User Activity Detection and Channel Estimation for Grant-Free Random Access in LEO Satellite-Enabled Internet-of-Things}

\author{Zhaoji~Zhang,~Ying~Li,~\IEEEmembership{Member,~IEEE,}~Chongwen~Huang,~\IEEEmembership{Member,~IEEE,}\\Qinghua Guo,~\IEEEmembership{Senior Member,~IEEE,}~Lei~Liu,~\IEEEmembership{Member,~IEEE,}~Chau Yuen,~\IEEEmembership{Senior Member,~IEEE,}\\and Yong Liang Guan,~\IEEEmembership{Senior Member,~IEEE}
\thanks{This work was presented in part at IEEE Global Communications Conference \cite{BGMP}. This work was supported in part by the National Natural Science Foundation of China (NSFC) under Grant 61971333, in part by the Key Industry Innovation Chain Project of Shaanxi under Grant 2018ZDCXLGY-04-04, in part by China Postdoctoral Science Foundation under Grant 2018M643580, and in part by A*STAR under its RIE2020 Advanced Manufacturing and Engineering (AME) Industry Alignment Fund – Pre Positioning (IAF-PP) (Grant No. A19D6a0053). Any opinions, findings and conclusions or recommendations expressed in this material are those of the author(s) and do not reflect the views of A*STAR. \emph{(Corresponding Author: Ying Li)}}
\thanks{Z. Zhang and Y. Li are with the State Key Laboratory of Integrated Services Network, Xidian University, Xi'an 710071, China (email: zjzhang\_1@stu.xidian.edu.cn; yli@mail.xidian.edu.cn).}
\thanks{C. Huang and C. Yuen are with the Engineering Product Development Pillar, Singapore University of Technology and Design, Singapore, 119613 (e-mail: chongwen\_huang@sutd.edu.sg; yuenchau@sutd.edu.sg).}
\thanks{Q. Guo is with the School of Electrical, Computer and Telecommunications Engineering, University of Wollongong, NSW 2522, Australia (e-mail: qguo@uow.edu.au).}
\thanks{L. Liu is with the School of Information Science, Japan Advanced Institute of Science and Technology (JAIST), 1-1 Asahidai, Nomi, Ishikawa 923-1292, Japan (e-mail: leiliu@jaist.ac.jp).}
\thanks{Y. L. Guan is with the School of Electrical and Electronic Engineering, Nanyang Technological University, Singapore 639798 (e-mail: eylguan@ ntu.edu.sg).}
}
\maketitle
\begin{abstract}
With recent advances on the dense low-earth orbit (LEO)  constellation, LEO satellite network has become one promising solution to providing global coverage for Internet-of-Things (IoT) services. Confronted with the sporadic transmission from randomly activated IoT devices, we consider the random access (RA) mechanism, and propose a grant-free RA (GF-RA) scheme to reduce the access delay to the mobile LEO satellites. A Bernoulli-Rician message passing with expectation maximization (BR-MP-EM) algorithm is proposed for this terrestrial-satellite GF-RA system to address the user activity detection (UAD) and channel estimation (CE) problem. This BR-MP-EM algorithm is divided into two stages. In the inner iterations, the Bernoulli messages and Rician messages are updated for the joint UAD and CE problem. Based on the output of the inner iterations, the expectation maximization (EM) method is employed in the outer iterations to update the hyper-parameters related to the channel impairments. Finally, simulation results show the UAD and CE accuracy of the proposed BR-MP-EM algorithm, as well as the robustness against the channel impairments.  
\end{abstract}
\begin{IEEEkeywords}
Low-earth orbit satellite, Internet-of-Things, grant-free random access, Bernoulli-Rician message passing, expectation maximization.
\end{IEEEkeywords}
\IEEEpeerreviewmaketitle
\section{Introduction}
\IEEEPARstart{T}{he} emerging Internet-of-Things (IoT) enables information exchange among physical objects, and enlightens the future development for a wide range of applications \cite{IoT1,IoT2,IoT3}, such as smart metering, e-health, fleet management, smart cities, etc. Therefore, supporting a variety of IoT applications becomes one major task for future wireless communication systems. Considering the engineering and operation cost, existing cellular communication networks are mainly deployed at populated areas, with the aim to meet the need for human-to-human (H2H) communications. However, compared with H2H devices, IoT devices are widely deployed in remote areas like deserts, coastal waters, and forests \cite{IoRT}. Therefore, it is challenging and cost-inefficient to provide services for remote IoT devices with existing cellular communication infrastructures.

As an alternative to cellular communication networks, the satellite communication network provides a promising solution to supporting IoT services \cite{satm2m}. Specifically, according to recent reports, thousands of low-earth orbit (LEO) satellites are planned to be launched by SpaceX \cite{spacex} and OneWeb \cite{oneweb}. Driven by these advances, a dense constellation of LEO satellites will be established to provide seamless Internet coverage for terrestrial users. Compared with existing cellular communication network and traditional geostationary earth orbit (GEO) satellite network, the LEO satellite network exhibits the following advantages in enabling IoT applications \cite{satiot}:

(i) Compared with existing cellular network, the terrestrial-satellite link (TSL) in LEO satellite communication is more robust in different terrestrial environment. Therefore, the LEO satellite network is more tolerant to geological disasters and extreme topographies like cliffs, valleys, and steep slopes.

(ii) Compared with cellular network, the LEO satellite network requires less support from terrestrial infrastructures such as base stations. Therefore, for IoT devices deployed in remote areas like deserts, forests and oceans, the LEO satellite network is more cost-efficient. More importantly, the LEO satellite network is also preferred for achieving global coverage due to its dense constellation.

(iii) The orbital altitude of LEO satellites (normally several hundred kilometers) is much lower than that of GEO satellites (approximately \SI{36000}{\kilo\metre}). Therefore, the propagation delay and path loss in LEO satellite communications are smaller than those in GEO satellite communications \cite{IoRT}. Furthermore, lower transmission power is required to reach LEO satellites than to reach GEO satellites, which is preferable for IoT devices with stringent requirements on power consumption. 

(iv) In GEO satellite communications, terrestrial devices can only access the satellite with fixed elevation angle. Therefore, GEO satellite communications is vulnerable to obstacles between the devices and the satellite receiver. However, in LEO satellite communications, IoT devices can access the mobile LEO satellites with flexible elevation angles. Therefore, LEO satellite-enabled IoT is tolerant to terrestrial obstacles.
\subsection{Related Works}\label{literature}
Motivated by above-mentioned advantages, some research has been conducted to exploit LEO satellites for future wireless communication network. For example, a dense LEO satellite access network (LEO-SAN) was proposed in \cite{Boya}, which integrates terrestrial communications with satellite communications. In this LEO-SAN, different physical-layer techniques such as interference management, diversity techniques, and cognitive radio schemes were investigated to achieve seamless coverage. A similar network architecture was proposed in \cite{Boya2} to integrate terrestrial-satellite network (TSN) into 5G and beyond to achieve efficient data offloading. The application of power-domain non-orthogonal multiple access (NOMA) was investigated in \cite{noma} for various satellite architectures. For IoT applications, an overview was presented in \cite{IoRT} on satellite-enabled Internet of Remote Things (IoRT), where various enabling techniques are discussed, such as the MAC protocol, resource allocation and transmission management. Furthermore, other topics including the constellation structure, spectrum allocation, and routing protocols are discussed in \cite{satiot}. The inter-plane inter-satellite link (ISL) was considered in \cite{matching} to further extend the coverage of IoT applications.

According to the service type \cite{servicetype, servicetype2}, IoT devices are intermittently activated with a certain probability and short data packets, i.e., active IoT devices perform random access (RA) for sporadic data transmission. Conventional RA schemes for satellite communications are mainly based on slotted ALOHA protocols \cite{IoRT,ALOHA1,ALOHA2}. In addition, a divide-and-conquer scheme was proposed in \cite{RA3} to allocate time slots to terminals based on service demands. Generally, these RA schemes assume sufficient access resources and static TSL, which becomes infeasible for IoT applications with massive connectivity and LEO satellites with high mobility.

Confronted with the massive connectivity in IoT and the rapidly-changing TSL to mobile LEO satellites, grant-free RA (GF-RA) schemes are preferred due to its spectral efficiency and low access delay. In GF-RA schemes, activated devices share the same access resource and directly transmit their data packets (along with pilot sequences), without applying for the grant from the satellite receiver. In this way, the signaling overhead can be reduced, which improves the transmission efficiency for the short data packets of IoT devices. 

Inspired by these advantages, a space diversity-based GF-RA scheme was proposed for satellite-enabled IoT \cite{GFRAsatellite}. However, one crucial task for the satellite receiver, i.e. the joint user activity detection (UAD) and channel estimation (CE) was bypassed in \cite{GFRAsatellite} by roughly modeling the TSL as an erasure-collision channel. To solve this joint UAD and CE problem, a typical solution is to formulate this task as a compressed sensing (CS) problem. In \cite{CS,CS1,CS2,CS3},  the pilot sequences from different devices serve as the sensing matrix in this CS problem, and the approximate message passing (AMP) algorithm was proposed for different system models. In addition, the variational Bayesian inference was employed in \cite{GQH,ZZJ,ZZJAPWCS}, where the UAD and CE are  solved by the mean-field message passing algorithm \cite{MF} and the Gaussian message passing algorithm \cite{GMPID,GMPCW,GMPCW2}, respectively.
\subsection{Motivations}
Existing RA schemes for satellite communications \cite{ALOHA1,ALOHA2,RA3} rely on static TSL and sufficient access resources. However, in LEO satellite-enabled IoT, the TSL is rapidly changing due to the mobility of the LEO satellite, and the massive connectivity from IoT devices may cause shortage of access resources. Therefore, existing RA schemes for satellite communications fail to match with the LEO satellite-enabled IoT. Although GF-RA schemes can improve the spectral efficiency and lower the transmission delay, the joint UAD and CE algorithms in existing GF-RA schemes \cite{CS,CS1,CS2,CS3,GQH,ZZJ,ZZJAPWCS} are designed for the Rayleigh channel, instead of the TSL channel. In the Rayleigh channel model, the line-of-sight (LoS) component in wave propagation is negligible, while the scattering component is dominant. However, both the LoS component and the scattering components should be considered for the TSL. To the best of our knowledge, there is little work tailored for the GF-RA scheme in LEO satellite-enabled IoT, which motivates our research in this paper.
\subsection{Contributions}
In this paper, we propose a terrestrial-satellite GF-RA scheme for the LEO satellite-enabled IoT. In order to address the joint UAD and CE problem, we propose a Bernoulli-Rician message passing with expectation maximization (BR-MP-EM) algorithm. The iterative message passing process of this BR-MP-EM algorithm is described on a factor graph and can be divided into two stages, i.e., the inner iterations and outer iterations. In the inner iterations, the Bernoulli messages and Rician messages are jointly updated for the UAD and CE. The channel impairments, i.e. the random propagation fading and phase shift, are treated as hyper-parameters. Then the EM method is employed in the outer iterations to update these hyper-parameters. The major contributions of this paper are listed as follows:

(i) A terrestrial-satellite GF-RA scheme is proposed for LEO satellite-enabled IoT, where active devices share the same access resource and directly transmit their data packets. This GF-RA scheme could avoid excessive signaling overhead, and thus improve the transmission efficiency for the short data packets of IoT devices.

(ii) With a typical channel model for the TSL, we derive the joint Bernoulli-Rician message passing in the inner iterations of the BR-MP-EM algorithm. The Bernoulli message update is derived for the UAD, while the Rician message update is derived for the CE.

(iii) In the outer iterations of the BR-MP-EM algorithm, the EM update is derived to estimate the channel impairment-related hyper-parameters, based on the output of the inner iterations. This EM update is independent from specific distributions of the hyper-parameters, and therefore robust against unknown channel impairments.
\subsection{Potential Applications}
A terrestrial-satellite GF-RA scheme is proposed in this paper, and we derive the BR-MP-EM algorithm to solve the joint UAD and CE problem in this GF-RA scheme. The proposed GF-RA scheme enables sporadic transmission from IoT devices to LEO satellites, and can be potentially applied to a wide range of IoT applications, especially when the IoT devices are remotely deployed so that the assistance of LEO satellites becomes essential. Some typical examples are smart grid, emergency management,
inshore and offshore windmills, land environment monitoring
and ocean monitoring \cite{IoRT}.
%to address the joint UAD and CE problem in the terrestrial-satellite GF-RA system. Instead of the Rayleigh channel, a widely-applied channel model \cite{LMS} for the LEO satellite was considered. In this BG-MP-EM algorithm, a factor graph with two sub-graphs is established for the message passing procedure, and the BG-MP-EM algorithm  divided into two stages, i.e., the inner iterations and outer iterations. The inner iterations are performed on the left sub-graph of the factor graph, where the Gaussian messages and Bernoulli messages \cite{AccessRA} are jointly updated. On the other hand, the outer iteration is performed on the right sub-graph, where the expectation maximization (EM) \cite{EM} method is employed to update the hyper-parameters related to the channel impairment. In this way, the Gaussian-Bernoulli message passing jointly works with the EM update to guarantee the UAD and CE accuracy, as well as the robustness against channel impairment.
\subsection{Section Organization}
This paper is organized as follows. In Section \ref{channel}, the system model is explained, and the joint UAD and CE problem is formulated for the terrestrial-satellite GF-RA system. In Section \ref{twouserana}, the BR-MP-EM algorithm is proposed, with detailed derivations on the Bernoulli-Rician message passing and the EM update. Simulation results are presented in Section \ref{capability} to evaluate the UAD and CE performance of the proposed BR-MP-EM algorithm, as well as its robustness against unknown channel impairments. Finally, Section \ref{conclusions} concludes this paper.
\section{System Model and Problem Formulation}\label{channel}
\subsection{System Model}\label{system_model_assum}
As illustrated in Fig. \ref{systemmodel}, we consider a terrestrial-satellite GF-RA system in LEO satellite-enabled IoT. Within the coverage area of a serving satellite, there are $K$ potential devices, each of which is randomly activated with a probability $p_a$. In each round of GF-RA, all the activated devices share the same access resource and transmit their data packets to the serving satellite (along with their unique pilot sequences). Considering the characteristics of dense LEO constellation and IoT applications, we adopt the following assumptions for this terrestrial-satellite GF-RA system:
\begin{figure}
	\centering
	\includegraphics[width=0.5\textwidth]{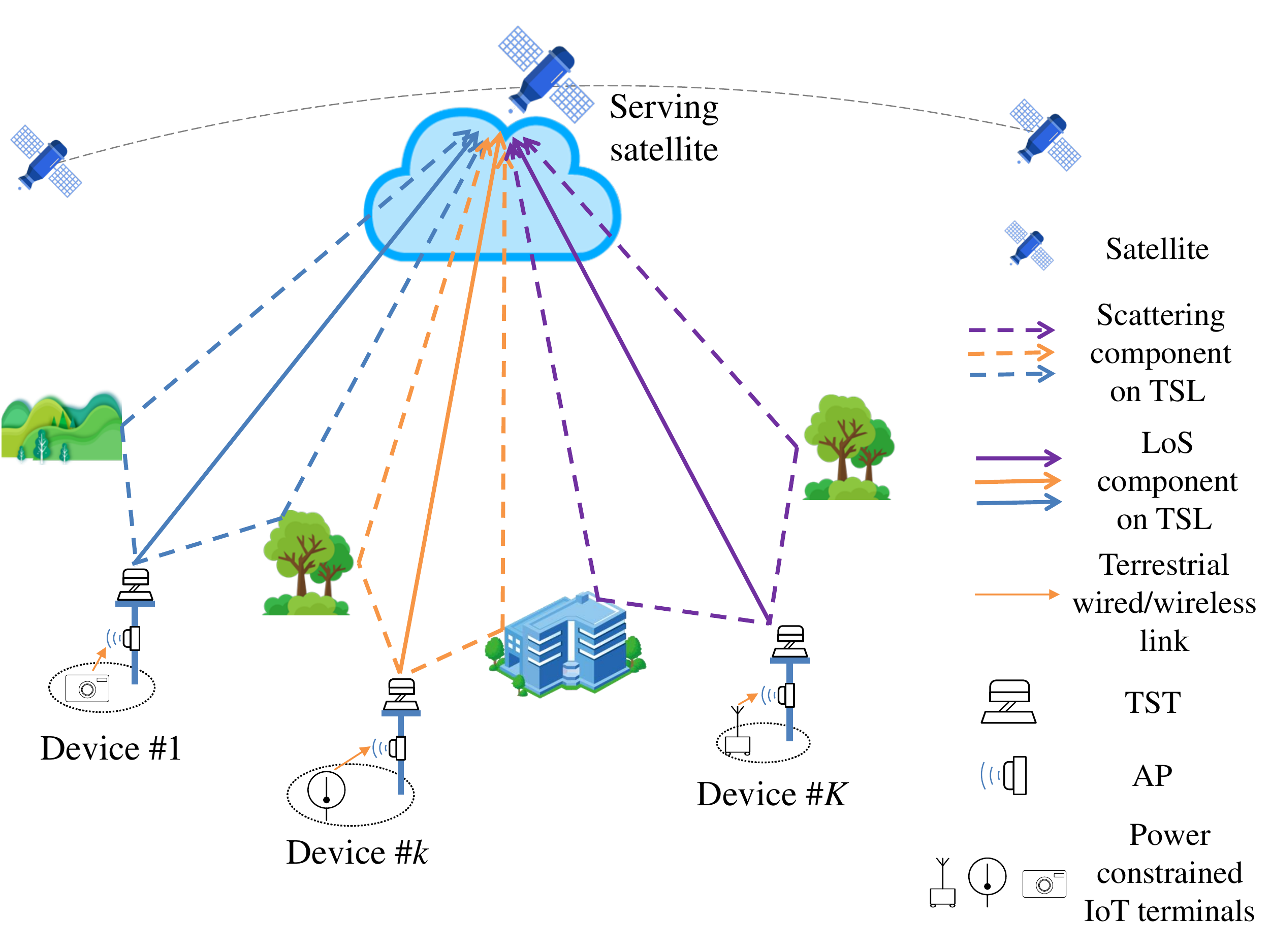}\vspace{-0.1cm}
	\caption{System model of the terrestrial-satellite GF-RA scheme in LEO satellite-enabled IoT.\vspace{-0.5cm}}\label{systemmodel}
\end{figure}

(i) Each device in this system refers to one physical access point (AP), which is equipped with a dedicated terrestrial-satellite terminal (TST) to facilitate transmission to the LEO satellite \cite{Boya,Boya2}. We assume that each AP provides coverage for a small number of nearby IoT terminals over wired or wireless terrestrial link. In this way, power-constrained sensors and IoT terminals can access the LEO satellite via the AP. Due to the low activity of these IoT terminals, the AP is also randomly activated in each round of GF-RA with a probability. 

(ii) We assume that all these $K$ devices are stationary, i.e., the devices are deployed at fixed locations with no mobility \cite{IoRT}. Furthermore, the surrounding terrestrial environment of these devices is assumed sufficiently static.

(iii) Since LEO satellites orbit the earth with fixed time period \cite{satiot}, we assume that aided by calibration from terrestrial control stations, the satellite receiver can maintain time synchronization with terrestrial devices. In addition, the timing advance information for each device's transmission is configured according to its location. In this way, data packets from different devices are aligned in time at the receiver.

(iv) On top of assumptions (ii) and (iii), we assume that activated devices perform GF-RA when the serving satellite appears at a fixed orbital position. In this way, in different rounds of GF-RA, each device can perform transmission at approximately fixed elevation angle.

For the TSL in this terrestrial-satellite GF-RA system, we adopt a widely-applied channel model, which is shown to match well with measured land mobile satellite (LMS) channel data \cite{LMS}. Specifically, the channel $h$ from each device to the LEO satellite is modeled as \cite{Boya,noma,LMS,LMS1,LMS2,LMS3}
\begin{equation}\label{LMSmodel}
{\color{red}h=h^\text{ray}e^{j\phi^\text{ray}}+h^\text{los}e^{j\phi^\text{los}},}
\end{equation}
where the combined effect of the scattering components between each device and the serving satellite (denoted by dashed lines in Fig. \ref{systemmodel}) is characterized by a Rayleigh-distributed amplitude $h^\text{ray}$ and a uniformly random phase $\phi^\text{ray}\sim\mathcal{U}[-\pi,\pi]$. On the other hand, the LoS component  between each device and the serving satellite (denoted by solid lines in Fig. \ref{systemmodel}) is characterized by a Nakagami-distributed amplitude $h^\text{los}$ and a deterministic phase $\phi^\text{los}$. It is noted that the random amplitude $h^\text{los}$ is mainly caused by the shadowing effect \cite{LMS}. 

On top of the LMS channel model (\ref{LMSmodel}), we add two additional factors to account for the channel impairments, i.e., the random propagation fading $h_r$ and  phase shift $\phi_\delta$, which will be explained later. Then, the terrestrial-satellite channel for each device $k$ is modeled as  
\begin{equation}\label{channelmodel}
h_k=h_re^{j\phi_{\delta}}\left(h^\text{ray}_ke^{j\phi^\text{ray}_k}+h^\text{los}_ke^{j\phi^\text{los}_k}\right).
\end{equation}
Compared with (\ref{LMSmodel}), a channel impairment term $h_re^{j\phi_{\delta}}$ is included in (\ref{channelmodel}), and we use subscript $k$ in the notations to distinguish the channel and wave-propagation components for different devices. It is noted that we can describe the scattering component $h^\text{ray}_ke^{j\phi^\text{ray}_k}$ in (\ref{channelmodel}) by a complex Gaussian distribution with variance $v^\text{ray}_k$, i.e., $h^\text{ray}_ke^{j\phi^\text{ray}_k}\sim\mathcal{CN}(0,v^\text{ray}_k)$. According to assumptions (ii) and (iv), the stationary devices perform GF-RA at fixed elevation angles. As a result, the shadowing effect on the LoS component is also sufficiently static in each round of GF-RA, so that the LoS amplitude $h^\text{los}_k$ can be considered as a constant in the proposed GF-RA scheme. In addition, we assume that all the devices experience the same phase shift $\phi_{\delta}$ and propagation fading $h_r$ in each round of GF-RA, and the reason for this assumption is explained in Remark \ref{commonfading}.
\subsection{Problem Formulation}
Assume that each device $k$ has a unique pilot sequence $\mathbf{P}_{k}=[P_{k1},\ldots,P_{kl},\ldots,P_{kL}]^T$ with length $L$. When device $k$ is activated, $\mathbf{P}_{k}$ is transmitted along with its data packet to facilitate the joint UAD and CE at the satellite receiver. Then, the $l$-th received pilot symbol $y_l$ is
\begin{equation}\label{symbolformulation}
y_l=\sum_{k\in\mathcal{K}}P_{kl}\alpha_kh_{k}+n_l,
\end{equation}
where $\mathcal{K}=\{1,\ldots K\}$ is the set of $K$ devices, $\alpha_k$ is the activity indicator, i.e., $\alpha_k=1$ if device $k$ is activated. Otherwise, $\alpha_k=0$. $h_k$ is the terrestrial-satellite channel of device $k$ in (\ref{channelmodel}), and $n_l$ is the additive white Gaussian noise (AWGN) with variance $\sigma_n^2$. Considering all the received pilot symbols, we have the following received pilot vector $\mathbf{y}$
\begin{equation}\label{matrixformulation}
\mathbf{y}_{L\times1}=\mathbf{P}_{L\times K}\left(\mathbf{h}_{K\times 1}\odot\mathbf{a}_{K\times 1}\right),
\end{equation}
where $\mathbf{y}_{L\times1}=[y_1,\ldots,y_l,\ldots,y_L]^T$ is the received pilot vector, $\mathbf{P}_{L\times K}=[\mathbf{P}_1,\ldots,\mathbf{P}_k,\ldots,\mathbf{P}_K]$ is the pilot matrix, $\mathbf{h}_{K\times 1}=[h_1,\ldots,h_k,\ldots,h_K]^T$ is the channel vector, $\mathbf{a}_{K\times 1}=[\alpha_1,\ldots,\alpha_k,\ldots,\alpha_K]^T$ is the activity vector, and $\odot$ refers to the element-wise multiplication. Therefore, this joint UAD and CE problem is formulated as follows
\begin{equation}\label{problemformulation}
\begin{split}
&\textbf{Solve:}\ \text{Detect}\ \alpha_k\ \text{and estimate}\ h_k \ \text{if}\ \alpha_k\ \text{is detected as}\ 1, \forall k\\
&\textbf{Given:}\ \mathbf{y}, \mathbf{P}, \overline{h}_r, h^\text{los}_k, \phi^\text{los}_k, v^\text{ray}_k,  \forall k 
\end{split}
\end{equation}
where $\overline{h}_r$ is the average propagation fading measured via statistical methods. According to assumptions (ii)-(iv), both the locations and transmission elevation angles of the stationary devices are fixed, which ensures the feasibility for the statistical measurement of $\overline{h}_r$ and other static parameters $h^\text{los}_k$, $\phi^\text{los}_k$, and $v^\text{ray}_k$.
\begin{remark}\label{commonfading} 
As in \cite{spacex,oneweb}, we employ Ku-band for the TSL in LEO satellite-enabled IoT. In the Ku-band, the propagation fading $h_r$ is related to the weather condition, the satellite receiver, and atmospheric variation \cite{modelreview}. We further assume that in each round of GF-RA, the LEO satellite only covers a relatively small area, where all $K$ devices experience similar weather condition due to geographical proximity. The assumption of geographical proximity can be justified by the fact that the coverage area of each LEO satellite will be greatly narrowed down in a forthcoming dense LEO constellation with thousands of satellites. In addition, each satellite can cover different small areas in a time-division manner, since IoT devices can be delay-tolerant in some typical applications \cite{IoRT,servicetype,servicetype2}. In this way, in each round of GF-RA, all $K$ devices experience the same propagation fading $h_r$ due to geographical proximity. On the other hand, $\phi_{\delta}$ models the phase shift caused by orbital perturbation, non-ideal elimination of Doppler shift \cite{DopplerElimination}, or atmospheric variation \cite{modelreview}. Similarly, we can assume that all $K$ devices experience the same $\phi_{\delta}$ due to geographical proximity.
\end{remark}
\begin{remark}\label{pilot}
In order to reduce the access delay for this GF-RA system, we adopt non-orthogonal Gaussian pilot sequences. Specifically, we assume that each pilot symbol is independently drawn from a complex Gaussian distribution, i.e., $P_{kl}\sim\mathcal{CN}(0,1), \forall k,l$. In contrast to orthogonal pilots with length $L=K$, the length of non-orthogonal Gaussian pilots can be chosen flexibly. Furthermore, we employ pilots with length $L<K$ to reduce the access delay and improve the transmission efficiency of short data packets.       
\end{remark}
\begin{figure}
	\centering
	\includegraphics[width=0.5\textwidth]{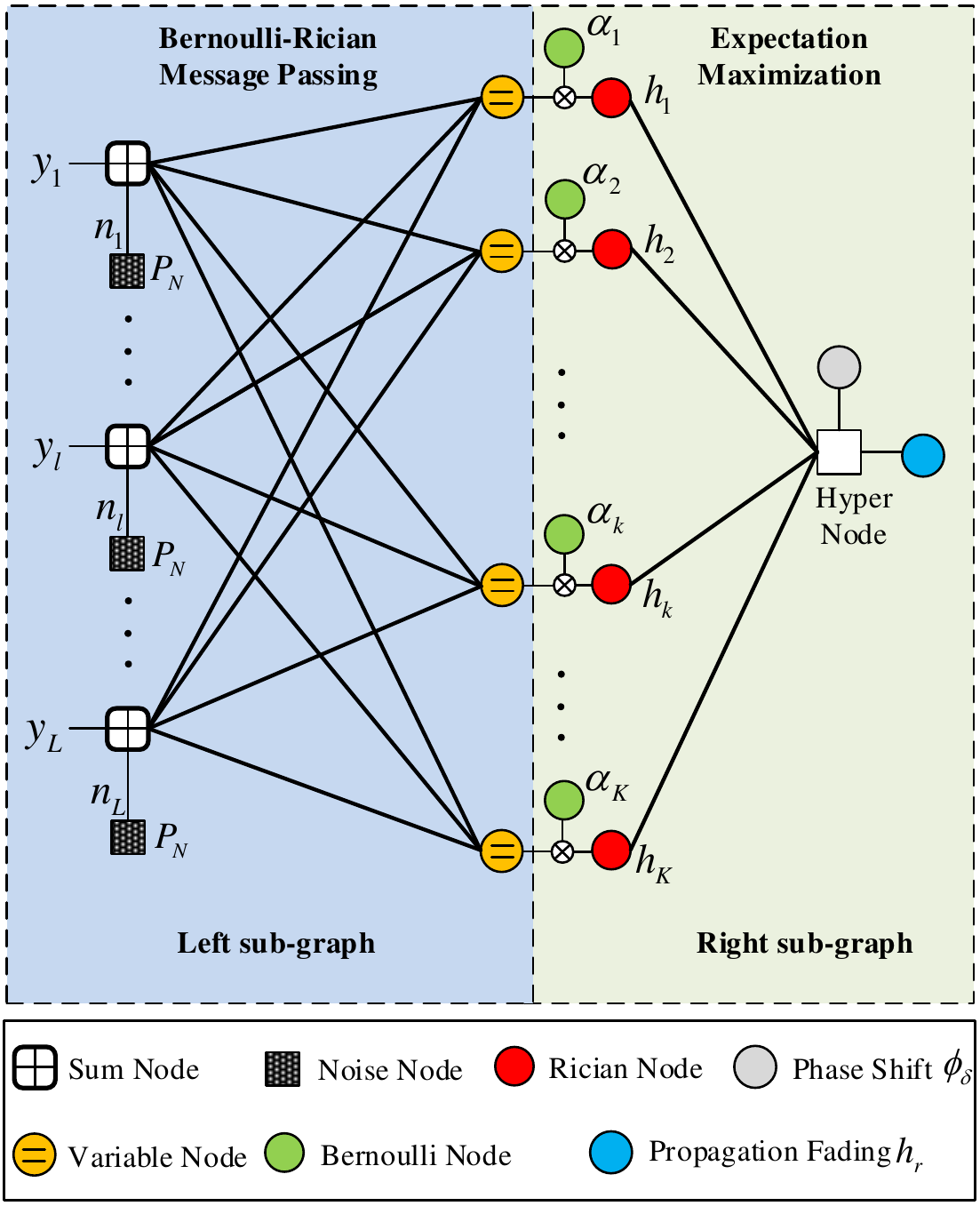}
	\caption{Factor graph of the BR-MP-EM algorithm. The inner iterations are performed on the left sub-graph, and the outer iteration is performed on the right sub-graph.}\label{factorgraph}
\end{figure}
\section{BR-MP-EM Algorithm for Joint UAD and CE}\label{twouserana}
In order to address the joint UAD and CE problem for this terrestrial-satellite GF-RA scheme, we derive the BR-MP-EM algorithm in this section. The message passing process of this BR-MP-EM algorithm is illustrated on the factor graph in Fig. \ref{factorgraph}, where the $l$-th sum node (SN) represents the summation term $\sum_{k\in\mathcal{K}}P_{kl}\alpha_kh_{k}$ in (\ref{symbolformulation}), the $k$-th variable node (VN) represents the product $\alpha_kh_{k}$ of the Bernoulli node $\alpha_k$ and the Rician node $h_k$ for device $k$, and the hyper-node represents the channel impairment term $h_re^{j\phi_{\delta}}$.
\begin{figure*}
	\centering
	\includegraphics[width=0.83\textwidth]{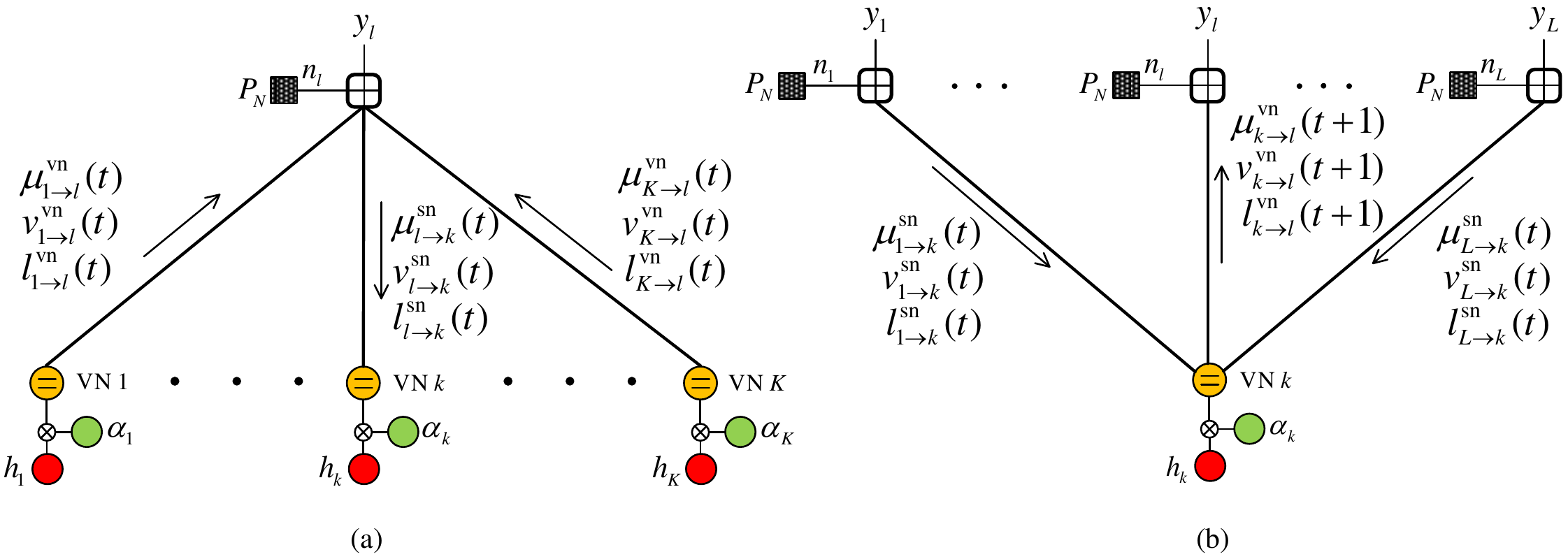}\vspace{-0.2cm}
	\caption{(a) Message update at sum nodes (SNs). (b) Message update at variable nodes (VNs).\vspace{-0.5cm}}\label{extrinsic}
\end{figure*}

As shown in (\ref{channelmodel}), the terrestrial-satellite channel $h_k$ of each device $k$ is a multiplication product of two terms, i.e. the Rician channel term and the channel impairment term $h_re^{j\phi_{\delta}}$. Accordingly, the message passing process on the factor graph is divided into two stages, i.e. the inner iterations (on the left sub-graph of Fig. \ref{factorgraph}) and the outer iterations (on the right sub-graph of Fig. \ref{factorgraph}). In the inner iterations, the propagation fading $h_r$ and phase shift $\phi_{\delta}$ are considered as known hyper-parameters, provided by the outer iterations. Then, we derive the Bernoulli-Rician message passing for the joint UAD and CE problem. In the outer iterations, the UAD and CE results from the left sub-graph are employed to update the estimates of hyper-parameters $\phi_{\delta}$ and $h_r$ with the EM method. Then another round of inner iterations is triggered by the updated hyper-parameters. The details of this BR-MP-EM algorithm are explained as follows.  
\subsection{Bernoulli-Rician Message Passing}
In this subsection, we assume that the hyper-parameters are fixed as $\hat{\phi}_{\delta}(\tau)$ and $\hat{h}_{r}(\tau)$ after the $\tau$-th outer iteration. As noted in \cite{viswanath} and in (\ref{channelmodel}), the Rician channel $h_k$ of each device $k$ contains a LoS component and a complex Gaussian scattering component. Furthermore, as discussed in Section \ref{system_model_assum}, the LoS component is assumed deterministic for our proposed terrestrial-satellite GF-RA system. Therefore, the Rician channel $h_k$ follows the complex Gaussian distribution  
\begin{equation}\label{pripdf}
h_k\sim\mathcal{CN}(\hat{h}_r(\tau)h^\text{los}_ke^{j(\phi^\text{los}_k+\hat{\phi}_{\delta}(\tau))}, \hat{h}^2_r(\tau)v^\text{ray}_k).
\end{equation}
In this way, the Rician message of the channel estimation can be characterized by the estimation mean-value $\mu_k$ and estimation variance $v_k$. In other words, $\mu_k$ is the estimate of $h_k$ in (\ref{pripdf}), while $v_k$ represents the deviation of this estimation. On the other hand, the Bernoulli message \cite{AccessRA} for the activity of device $k$ is represented by a probability $p_k$. That is, $p_k$ is the probability for $\alpha_k$ to take the value 1. Then, we can derive the Bernoulli-Rician message passing for the joint UAD and CE problem on the left sub-graph of Fig. \ref{factorgraph}. The message passing diagram is illustrated in Fig. \ref{extrinsic}.
\subsubsection{Message Update at Sum Nodes}
Denote $p^\text{vn}_{k\to l}(t)$ as the Bernoulli message for the activity of device $k$, which is  passed from VN $k$ to SN $l$ in the $t$-th inner iteration. Denote $\mu^\text{vn}_{k\to l}(t)$ and $v^\text{vn}_{k\to l}(t)$ as the Rician messages passed from VN $k$ to SN $l$. That is, $\mu^\text{vn}_{k\to l}(t)$ and $v^\text{vn}_{k\to l}(t)$ represent the estimate and estimation deviation of the channel $h_k$, respectively. Then we rewrite (\ref{symbolformulation}) as follows,
\begin{equation}\label{equinoise}
y_l=P_{kl}\alpha_kh_{k}+\underbrace{\sum_{j\in\mathcal{K}/ k}P_{jl}\alpha_jh_{j}+n_l}_{n^*_{lk}},
\end{equation}
where $\sum\limits_{j\in\mathcal{K}/ k}P_{jl}\alpha_jh_{j}+n_l$ is approximated as an equivalent noise $n^*_{lk}\sim\mathcal{CN}\left(\mu^*_{lk}(t),v^*_{lk}(t)\right)$. Based on the incoming Bernoulli messages and Rician messages, we can derive $\mu^*_{lk}(t)$ and $v^*_{lk}(t)$ as follows,
\begin{equation}\label{equimuvar}
\begin{split}
\mu^*_{lk}(t)&\!=\!\sum_{j\in\mathcal{K}/ k}P_{jl}p^\text{vn}_{j\to l}(t)\mu^\text{vn}_{j\to l}(t)\\
v^*_{lk}(t)&\!=\!\sigma^2_n\!\!+\!\!\!\sum_{j\in\mathcal{K}/ k}\!\!|P_{jl}|^2p^\text{vn}_{j\to l}(t)\!\left[v^\text{vn}_{j\to l}(t)\!+\!q^\text{vn}_{j\to l}(t)|\mu^\text{vn}_{j\to l}(t)|^2\right]
\end{split}
\end{equation}
where $\sigma_n^2$ represents the noise variance of $n_l$ in (\ref{equinoise}). $q^\text{vn}_{j\to l}(t)$ represents the probability for the Bernoulli variable $\alpha_j$ to take the value $0$, i.e., $q^\text{vn}_{j\to l}(t)=1-p^\text{vn}_{j\to l}(t)$. Then the messages passed from SN $l$ to VN $k$ are derived as follows.

\noindent\textbf{\underline{Rician Message Update at SN}:} The Rician messages of $h_k$, i.e. the estimate $\mu^\text{sn}_{l\to k}(t)$ and estimation deviation $v^\text{sn}_{l\to k}(t)$ passed from SN $l$ to VN $k$ in the $t$-th inner iteration are derived as follows
\begin{equation}\label{SNupdateGaussian}
\begin{split}
\mu^\text{sn}_{l\to k}(t){=}&\mathbf{E}\left[h_k|y_l,\mu^*_{lk}(t),v^*_{lk}(t),\alpha_k=1\right]\\
     =&(y_l-\mu^*_{lk}(t))/P_{kl}\\
v^\text{sn}_{l\to k}(t){=}&\mathbf{Var}\left[h_k|y_l,\mu^*_{lk}(t),v^*_{lk}(t),\alpha_k=1\right]\\
=&v^*_{lk}(t)/|P_{kl}|^2
\end{split}
\end{equation}
where $\mathbf{E}\left[a|b\right]$ and $\mathbf{Var}\left[a|b\right]$ represent the expectation and variance of $a$ conditioned on $b$, respectively.

\noindent\textbf{\underline{Bernoulli Message Update at SN}:} The Bernoulli message of $\alpha_k$, i.e. the non-zero probability $p^\text{sn}_{l\to k}(t)$ passed from SN $l$ to VN $k$ in the $t$-th inner iteration is derived as follows
\begin{equation}\label{SNupdateBernoulli}
\begin{split}
p^\text{sn}_{l\to k}(t)&=\left[1+\frac{\mathbb{P}\left(y_l|\alpha_k=0,\mu^*_{lk}(t),v^*_{lk}(t)\right)}{\mathbb{P}\left(y_l|\alpha_k=1,\mu^*_{lk}(t),v^*_{lk}(t)\right)}\right]^{-1}\\
&=\left[1+\frac{\mathbb{P}\left(y_l=n^*_{lk}|\mu^*_{lk}(t),v^*_{lk}(t)\right)}{\mathbb{P}\left(y_l=P_{kl}h_k+n^*_{lk}|\mu^*_{lk}(t),v^*_{lk}(t)\right)}\right]^{-1}\\
&=\left[1+\frac{f\left(y_l|\mu^*_{lk}(t),v^*_{lk}(t)\right)}{f\left(y_l|\mu^1_{lk}(t),v^1_{lk}(t)\right)}\right]^{-1}
\end{split}
\end{equation}
where $\mu^1_{lk}(t)=P_{kl}\mu^\text{vn}_{k\to l}(t)+\mu^*_{lk}(t)$ and $v^1_{lk}(t)=v^*_{lk}(t)+|P_{kl}|^2v^\text{vn}_{k\to l}(t)$ represent the mean-value and variance of $y_l$ when $\alpha_k=1$, and $f(x|\mu,\sigma^2)$ represents the probability density function (pdf) of a complex Gaussian distribution $\mathcal{CN}\left(\mu,\sigma^2\right)$, i.e.,
\begin{equation}\label{pdf}
f(x|\mu,\sigma^2)=\frac{1}{\pi \sigma^2}\exp\left(-\frac{|x-\mu|^2}{\sigma^2}\right).
\end{equation}

In order to avoid the computation overflow caused by a large number of multiplications of probabilities, we employ the log-likelihood ratio (LLR) to represent the Bernoulli message. The relationship between the LLR message $l^\text{sn}_{l\to k}(t)$ and the non-zero probability $p^\text{sn}_{l\to k}(t)$ is defined as
\begin{equation}\label{LLR}
\begin{split}
l^\text{sn}_{l\to k}(t)\overset{\Delta}{=}&\ln\frac{p^\text{sn}_{l\to k}(t)}{1-p^\text{sn}_{l\to k}(t)},\\
\overset{(a)}{=}&\ln\frac{v^*_{lk}(t)}{v^1_{lk}(t)}+\frac{|y_l-\mu^*_{lk}(t)|^2}{v^*_{lk}(t)}-\frac{|y_l-\mu^1_{lk}(t)|^2}{v^1_{lk}(t)},\\
p^\text{sn}_{l\to k}(t)\overset{\Delta}{=}&\frac{1}{1+\exp\left(-l^\text{sn}_{l\to k}(t)\right)},
\end{split}
\end{equation}
where equation ($a$) of (\ref{LLR}) is derived by substitution of (\ref{SNupdateBernoulli}). 
\subsubsection{Message Update at Variable Nodes}
The messages passed from VNs to SNs follow the message combination rule \cite{messagecombination}. The Rician messages and Bernoulli messages are updated as follows.

\noindent\textbf{\underline{Rician Message Update at VN}:} For the Rician messages, the estimate $\mu^\text{vn}_{k\to l}(t+1)$ and estimation deviation
$v^\text{vn}_{k\to l}(t+1)$ of $h_k$ passed from VN $k$ to SN $l$ are derived as follows,
\begin{equation}\label{VNupdateGaussian}
\begin{split}
v^\text{vn}_{k\to l}(t+1)&=\mathbf{Var}\left[h_k|\hat{h}_{r}(\tau),v^\text{ray}_k,\mathbf{v}^\text{sn}_{\mathcal{L}/l}(t)\right]\\
&=\left[\left(v^\text{pri}_k(\tau)\right)^{-1}+\sum_{i\in\mathcal{L}/l}\left(v^\text{sn}_{i\to k}(t)\right)^{-1}\right]^{-1}\\
\mu^\text{vn}_{k\to l}(t+1)&=\mathbf{E}\left[h_k|\hat{\phi}_{\delta}(\tau),\hat{h}_{r}(\tau),v^\text{ray}_k,\mathbf{\mu}^\text{sn}_{\mathcal{L}/l}(t)\right]\\
&=v^\text{vn}_{k\to l}(t+1)\left[ \frac{\mu^\text{pri}_k(\tau)}{v^\text{pri}_k(\tau)} + \sum_{i\in\mathcal{L}/l}\frac{\mu^\text{sn}_{i\to k}(t)}{v^\text{sn}_{i\to k}(t)} \right]
\end{split}
\end{equation}
where $\mu^\text{pri}_k(\tau)=\hat{h}_{r}(\tau)h^\text{los}_ke^{j(\phi^\text{los}_k+\hat{\phi}_{\delta}(\tau))}$ and $v^\text{pri}_k(\tau)=\hat{h}^2_{r}(\tau)v^\text{ray}_k$ represent the prior mean-value and variance of $h_k$ after the $\tau$-th outer iteration, $\mathcal{L}=\{1,\ldots,L\}$ is the set of all the SNs, $\mathbf{\mu}^\text{sn}_{\mathcal{L}/l}(t)=\{\mu^\text{sn}_{i\to k}(t), \forall i\in\mathcal{L}/l\}$, and $\mathbf{v}^\text{sn}_{\mathcal{L}/l}(t)=\{v^\text{sn}_{i\to k}(t), \forall i\in\mathcal{L}/l\}$.

\noindent\textbf{\underline{Bernoulli Message Update at VN}:} The Bernoulli message $p^\text{vn}_{k\to l}(t+1)$ passed from VN $k$ to SN $l$  is derived as follows,
\begin{equation}\label{VNupdateBernoulli}
\begin{split}
p^\text{vn}_{k\to l}(t+1)&\!=\!\mathbb{P}\left(\alpha_k=1|\mathbf{p}^\text{sn}_{\mathcal{L}/l}(t),p_a\right)\\
&\!=\!\frac{p_a\prod\limits_{i\in\mathcal{L}/l}p^\text{sn}_{i\to k}(t)}{p_a\prod\limits_{i\in\mathcal{L}/l}p^\text{sn}_{i\to k}(t)+(1\!-\!p_a)\!\!\prod\limits_{i\in\mathcal{L}/l}(1-p^\text{sn}_{i\to k}(t))}
\end{split}
\end{equation}
where $\mathbf{p}^\text{sn}_{\mathcal{L}/l}(t)=\{p^\text{sn}_{i\to k}(t), \forall i\in\mathcal{L}/l\}$, and $p_a$ is the prior activation probability of each device. Again, for computational convenience, we derive the LLR message $l^\text{vn}_{k\to l}(t+1)$ for the Bernoulli variable $\alpha_k$,
\begin{equation}\label{VNupdateBernoulliLLR}
l^\text{vn}_{k\to l}(t+1)\overset{\Delta}{=}\ln\frac{p^\text{vn}_{k\to l}(t+1)}{1-p^\text{vn}_{k\to l}(t+1)}=l_0+\sum_{i\in\mathcal{L}/l}l^\text{sn}_{i\to k}(t),
\end{equation}
where $l_0=\ln\frac{p_a}{1-p_a}$ is the prior LLR for the activity of each device. Note that, in the calculation of $\mu^*_{lk}(t+1)$ and $v^*_{lk}(t+1)$ in (\ref{equimuvar}), the non-zero probability $p^\text{vn}_{k\to l}(t+1)$ in (\ref{VNupdateBernoulli}) will be used for the $(t+1)$-th SN update, and we have
\begin{equation}\label{reverseLLRtrans}
p^\text{vn}_{k\to l}(t+1)=\frac{1}{1+\exp\left({-l^\text{vn}_{k\to l}(t+1)}\right)}.
\end{equation}
\subsubsection{CE Output and UAD Decision at Variable Nodes} After a fixed number (denoted by $N_{in}$) of inner iterations, each VN outputs its final CE and UAD decision. Similar to the message update at VNs, the CE output and UAD decision are obtained by the message combination rule over all the incoming messages.

\noindent\underline{\textbf{CE Output}:} The final channel estimate $\mu^\text{dec}_k(t)$ and estimation deviation $v^\text{dec}_k(t)$ for device $k$ are obtained by combining all the incoming Rician messages about $h_k$, i.e.,  
\begin{equation}\label{VNCE}
\begin{split}
v^\text{dec}_{k}(t)&=\left[\left(v^\text{pri}_k(\tau)\right)^{-1}+\sum_{l\in\mathcal{L}}\left(v^\text{sn}_{l\to k}(t)\right)^{-1}\right]^{-1},\\
\mu^\text{dec}_{k}(t)&=v^\text{dec}_{k}(t)\left[ \frac{\mu^\text{pri}_k(\tau)}{v^\text{pri}_k(\tau)} + \sum_{l\in\mathcal{L}}\frac{\mu^\text{sn}_{l\to k}(t)}{v^\text{sn}_{l\to k}(t)} \right].
\end{split}
\end{equation}
\noindent\underline{\textbf{UAD Decision}:} The LLR for UAD decision is expressed as
\begin{equation}\label{VNUAD}
l^\text{dec}_{k}(t)=l_0+\sum_{l\in\mathcal{L}}l^\text{sn}_{l\to k}(t)+l^\text{ce}_k(t).
\end{equation}
If $l^\text{dec}_{k}(t)>0$, the activity is detected as $\hat{\alpha}_k(t)=1$, and then $\mu^\text{dec}_{k}(t)$ in (\ref{VNCE}) is considered as estimated channel for device $k$. Otherwise, $\hat{\alpha}_k(t)=0$. Note that compared with (\ref{VNupdateBernoulliLLR}), an additional LLR term $l^\text{ce}_k(t)$ is included in (\ref{VNUAD}) to exploit the CE output to improve the UAD accuracy. The motivation to include $l^\text{ce}_k(t)$ and its derivation are detailed as follows.

\noindent\underline{\textbf{Derivation of} $l^\text{ce}_k(t)$}\textbf{:} Note that if the final estimate $\mu^\text{dec}_{k}(t)$ in (\ref{VNCE}) is close to $0$, it is more likely that this device is inactive. Therefore, we can exploit the CE result in (\ref{VNCE}) to improve the UAD accuracy. The estimated channel $\hat{h}_k$ can be expressed as $\hat{h}_k=h_k+e_k$, where $e_k$ represents the estimation error caused by channel noise and multi-user interference. Since $v^\text{dec}_{k}(t)$ in (\ref{VNCE}) represents the CE deviation, we approximate $e_k$ as a complex Gaussian random variable with variance $v^\text{dec}_{k}(t)$, i.e. $e_k\sim\mathcal{CN}\left(0,v^\text{dec}_{k}(t)\right)$. Note that if $\alpha_k=1$, the prior information is that  $h_k\sim\mathcal{CN}(\mu^\text{pri}_k(\tau),v^\text{pri}_k(\tau))$, and we further have $\hat{h}_k\sim\mathcal{CN}(\mu^\text{pri}_k(\tau),v^\text{pri}_k(\tau)+v^\text{dec}_{k}(t))$. On the other hand, $h_k$ is equivalently equal to $0$ if $\alpha_k=0$, and then we have $\hat{h}_k\sim\mathcal{CN}(0,v^\text{dec}_{k}(t))$. Therefore, $l^\text{ce}_k(t)$ is derived as follows
\begin{equation}\label{ceforllr}
\begin{split}
l^\text{ce}_k(t)&\!=\!\ln\frac{\mathbb{P}(\hat{h}_k=\mu^\text{dec}_{k}(t)|\alpha_k=1,\mu^\text{pri}_k(\tau),v^\text{pri}_k(\tau),v^\text{dec}_{k}(t))}{\mathbb{P}(\hat{h}_k=\mu^\text{dec}_{k}(t)|\alpha_k=0,v^\text{dec}_{k}(t))}\\
&\!=\!\ln\frac{f(\mu^\text{dec}_{k}(t)|\mu^\text{pri}_k(\tau),v^\text{pri}_k(\tau)+v^\text{dec}_{k}(t))}{f(\mu^\text{dec}_{k}(t)|0,v^\text{dec}_{k}(t))}\\
&\!=\!\ln\frac{v^\text{dec}_{k}(t)}{v^\text{pri}_k(\tau)\!+\!v^\text{dec}_{k}(t)}+\frac{|\mu^\text{dec}_{k}(t)|^2}{v^\text{dec}_{k}(t)}-\frac{|\mu^\text{dec}_{k}(t)\!-\!\mu^\text{pri}_k(\tau)|^2}{v^\text{pri}_k(\tau)\!+\!v^\text{dec}_{k}(t)}
\end{split}
\end{equation}
where  $f(x|\mu,\sigma^2)$ is defined in (\ref{pdf}).
\subsection{Expectation-Maximization Update of Hyper-Parameters}
If device $k$ is detected as active after a certain number of inner iterations, $\mu^\text{dec}_{k}(t)$ and $v^\text{dec}_{k}(t)$ are passed to the right sub-graph of Fig. \ref{factorgraph} to update the hyper-parameters $h_r$ and $\phi_{\delta}$ with the EM method. Note that the Rician messages  $\mu^\text{dec}_{k}(t)$ and $v^\text{dec}_{k}(t)$ from the left sub-graph of Fig. \ref{factorgraph} characterize the posterior complex Gaussian pdf  $f(h_k|\mu^\text{dec}_{k}(t), v^\text{dec}_{k}(t);\mathbf{y},\hat{h}_{r}(\tau),\hat{\phi}_{\delta}(\tau))$ of $h_k$. For notational simplicity, we rewrite this posterior Gaussian pdf as $f^\text{pos}_\tau(h_k)$,
\begin{equation}
f^\text{pos}_\tau(h_k)\overset{\Delta}{=}f(h_k|\mu^\text{dec}_{k}(t), v^\text{dec}_{k}(t);\mathbf{y},\hat{h}_{r}(\tau),\hat{\phi}_{\delta}(\tau)).
\end{equation}
It is shown that $f^\text{pos}_\tau(h_k)$ is obtained with given hyper-parameters $\hat{h}_{r}(\tau)$ and $\hat{\phi}_{\delta}(\tau)$ after the $\tau$-th outer iteration. According to (\ref{pripdf}), we define the prior Gaussian pdf $f^\text{pri}(h_k;h_r,\phi_{\delta})$ conditioned on hyper-parameters $h_r$ and $\phi_{\delta}$,
\begin{equation}\label{priGaussianpdf}
f^\text{pri}(h_k;h_r,\phi_{\delta})\overset{\Delta}{=}f(h_k|h_rh^\text{los}_ke^{j(\phi^\text{los}_k+\phi_{\delta})}, h^2_rv^\text{ray}_k)
\end{equation}
Then, the EM update in the ($\tau+1$)-th outer iteration aims to find a pair of hyper-parameters $\hat{h}_{r}(\tau+1)$ and $\hat{\phi}_{\delta}(\tau+1)$ that maximizes the expectation $\mathbf{E}\left[f^\text{pri}(h_k;h_r,\phi_{\delta})\right]$ over the distribution $f^\text{pos}_\tau(h_k)$, i.e.,
\begin{equation}\label{EMoneuser}
\begin{split}
&\{\hat{h}_{r}(\tau+1), \hat{\phi}_{\delta}(\tau+1)\}\\
&=\mathop{\arg\max}_{h_r,\phi_\delta}\ \mathbf{E}\left[f^\text{pri}(h_k;h_r,\phi_{\delta})\right]\\
&\overset{(b)}{=}\mathop{\arg\max}_{h_r,\phi_\delta}\ \mathbf{E}\left[\ln f^\text{pri}(h_k;h_r,\phi_{\delta})\right]\\
&\overset{(c)}{=}\mathop{\arg\max}_{h_r,\phi_\delta}\ \int{f^\text{pos}_\tau(h_k)\ln f^\text{pri}(h_k;h_r,\phi_{\delta})}dh_k\\
\end{split}
\end{equation}
where equation ($b$) of (\ref{EMoneuser}) is obtained by the fact that $\ln(x)$ is a monotonic function, and equation ($c$) of (\ref{EMoneuser}) is obtained by the fact that the expectation $\mathbf{E}[\cdot]$ is taken over $h_k$ with posterior pdf $f^\text{pos}_\tau(h_k)$.

Considering that all the active devices share the same hyper-parameters, we further extend the EM update equation (\ref{EMoneuser}) from the single-user case to the multi-user case, i.e., 
\begin{equation}\label{EMmoreuser}
\begin{split}
&\{\hat{h}_{r}(\tau+1), \hat{\phi}_{\delta}(\tau+1)\}\\
&=\mathop{\arg\max}_{h_r,\phi_\delta}\ \mathbf{E}\left[\sum\limits_{k\in\mathcal{K}^+}\ln f^\text{pri}(h_k;h_r,\phi_{\delta})\right]\\
&\!=\!\mathop{\arg\max}_{h_r,\phi_\delta}\!\!\int\!\!\!\ldots\!\!\!\int\!\!\!{\prod\limits_{k\in\mathcal{K}^+}\!\!\!f^\text{pos}_\tau(\!h_k\!)\!\!\left(\sum\limits_{k\in\mathcal{K}^+}\!\!\ln f^\text{pri}(\!h_k;\!h_r,\phi_{\delta})\!\!\right)}\!\!\prod\limits_{k\in\mathcal{K}^+}\!\!\!dh_k\\
\end{split}
\end{equation}
where $\mathcal{K^+}$ is the set of devices that are detected as active and participate in the EM update. To solve the maximization problem in (\ref{EMmoreuser}), we differentiate $\mathbf{E}\left[\sum\limits_{k\in\mathcal{K}^+}\ln f^\text{pri}(h_k;h_r,\phi_{\delta})\right]$ with respect to $h_r$ and $\phi_\delta$ respectively, and set both partial derivatives to $0$. In this way, we have the following set of equations,
\begin{equation}\label{equationset}
\left\{ \begin{split}
\int\!\!\!\ldots\!\!\!\int\!\!\!{\prod\limits_{k\in\mathcal{K}^+}\!\!\!f^\text{pos}_\tau(\!h_k\!)\!\!\left(\sum\limits_{k\in\mathcal{K}^+}\!\!\frac{\partial\ln f^\text{pri}(\!h_k;\!h_r,\phi_{\delta})}{\partial h_r} \!\!\right)}\!\!\prod\limits_{k\in\mathcal{K}^+}\!\!\!dh_k=0,\\
\int\!\!\!\ldots\!\!\!\int\!\!\!{\prod\limits_{k\in\mathcal{K}^+}\!\!\!f^\text{pos}_\tau(\!h_k\!)\!\!\left(\sum\limits_{k\in\mathcal{K}^+}\!\!\frac{\partial\ln f^\text{pri}(\!h_k;\!h_r,\phi_{\delta})}{\partial \phi_{\delta}} \!\!\right)}\!\!\prod\limits_{k\in\mathcal{K}^+}\!\!\!dh_k=0. 
\end{split}
\right.
\end{equation}
By solving (\ref{equationset}), we have the solution to (\ref{EMmoreuser}). Then, the hyper-parameters in the ($\tau+1$)-th outer iteration are updated as follows
\begin{equation}\label{solution}
\left\{ \begin{split}
\hat{\phi}_\delta(\tau+1)&=\angle M(\tau)\\
\hat{h}_r(\tau+1)&=\frac{-|M(\tau)|+\sqrt{|M(\tau)|^2+4N(\tau)}}{2} 
\end{split}
\right.
\end{equation} 
where $\angle M(\tau)$ represents the angle of a complex number $M(\tau)$, $M(\tau)=\left\langle\left(\mu^\text{dec}_{k}(t)h^\text{los}_ke^{-j\phi^\text{los}_k}\right)/{v^\text{ray}_k}\right\rangle$, $N(\tau)=\left\langle\left(v^\text{dec}_{k}(t)+|\mu^\text{dec}_{k}(t)|^2\right)/{v^\text{ray}_k}\right\rangle$. $\left\langle \cdot\right\rangle$ represents the averaging operation, i.e., $\left\langle x_k\right\rangle=\sum\limits_{k\in\mathcal{K^+}}x_k/\|\mathcal{K^+}\|$ where $\|\mathcal{K^+}\|$ is the number of elements in the set $\mathcal{K^+}$. The detailed derivation of (\ref{solution}) is explained in Appendix \ref{derivation}.
\begin{algorithm}[t]
	\setstretch{1.2}
	\caption{BR-MP-EM Algorithm}
	\label{alg:BRMPEM}
	\textbf{Initialization:} 
	
	$\tau=1$, $\hat{h}_r(\tau)=\overline{h}_r$, $\hat{\phi}_\delta(\tau)=0$, $t=1$, $v^\text{vn}_{k\to l}(t)=\hat{h}^2_r(\tau)v^\text{ray}_k$, $\mu^\text{vn}_{k\to l}(t)=\hat{h}_r(\tau)e^{j\hat{\phi}_\delta(\tau)} h^\text{los}_ke^{j\phi^\text{los}_k}$, $l^\text{vn}_{k\to l}(t)=\ln\frac{p_a}{1-p_a}$.
	
	1:\textbf{for} $\tau=1:1:N_{out}$
	
	2: \ \ \ \textbf{for} $t=1:1:N_{in}$
	
	3: \ \ \ \ \ SN update $\mu^\text{sn}_{l\to k}(t)$, $v^\text{sn}_{l\to k}(t)$ by (\ref{SNupdateGaussian}).
	
	4: \ \ \ \ \ SN update $l^\text{sn}_{l\to k}(t)$ by (\ref{LLR}).
	
	5: \ \ \ \ \ VN update $\mu^\text{vn}_{k\to l}(t+1)$, $v^\text{vn}_{k\to l}(t+1)$ by (\ref{VNupdateGaussian}).
	
	6: \ \ \ \ \ VN update $l^\text{vn}_{k\to l}(t+1)$ by (\ref{VNupdateBernoulliLLR}).
	
	7: \ \ \ \textbf{end}
	
	8: CE output $\mu^\text{dec}_{k}(N_{in})$, $\mu^\text{dec}_{k}(N_{in}-1)$, and  $v^\text{dec}_{k}(N_{in})$ by (\ref{VNCE}).
	
	9: UAD decision $l^\text{dec}_{k}(N_{in})$ by (\ref{VNUAD}).
	
	10:Relative variation $\Delta_k=\frac{|\mu^\text{dec}_{k}(N_{in})-\mu^\text{dec}_{k}(N_{in}-1)|}{|\mu^\text{dec}_{k}(N_{in}-1)|}$.
	
	11:EM index $\mathcal{K}^+=\left\{l^\text{dec}_{k}(N_{in})>0 \ \text{and}\ \Delta_k<\eta_{th}, \forall k\right\}$
	
	12:EM update of $\hat{h}_{r}(\tau+1), \hat{\phi}_{\delta}(\tau+1)$ by (\ref{solution})
	
	13:$\mu^\text{vn}_{k\to l}(1)=\mu^\text{vn}_{k\to l}(N_{in}+1), v^\text{vn}_{k\to l}(1)=v^\text{vn}_{k\to l}(N_{in}+1)$,
	
	14:$l^\text{vn}_{k\to l}(1)=l^\text{vn}_{k\to l}(N_{in}+1)$.
	
	15:\textbf{end}
	
	\textbf{Final Decision:}
	
	$\hat{\mathbf{h}}\!\!=\!\!\left\{\hat{\alpha}_k\mu^\text{dec}_{k}(N_{in}),\forall k\right\}, \hat{\mathbf{a}}\!\!=\!\!\left\{\hat{\alpha}_k,\forall k\right\}$ with $\hat{\alpha}_k$ obtained by (\ref{VNUAD}).
\end{algorithm}
\subsection{Summary of BR-MP-EM Algorithm}
The proposed BR-MP-EM algorithm is summarized in Algorithm \ref{alg:BRMPEM}. In Algorithm \ref{alg:BRMPEM}, $N_{out}$ is the number of outer iterations, and $N_{in}$ is the number of inner iterations. That is, a total number of $N_{it}=N_{out}N_{in}$ iterations are performed by the BR-MP-EM algorithm. Specifically, Line 3-4 represent the Rician message update and Bernoulli message update at SNs, respectively. Line 5-6 represent the Rician message update and Bernoulli message update at VNs, respectively. Line 8-9 represent the CE output and UAD decision of the inner iterations, respectively. Note that in Line 10, we calculate the relative variation $\Delta_k$  to evaluate the convergence of the channel estimation. In line 11, we choose the set of devices for the EM update in the outer iteration. That is, the channel estimate of device $k$ will be included in the EM update if device $k$ is detected as active and the relative variation $\Delta_k$ is smaller than a threshold $\eta_{th}$. It is noted that the threshold $\eta_{th}$ is considered to avoid error propagation from inner iterations to outer iterations, and the value of $\eta_{th}$ ($0<\eta_{th}\leq1$) is chosen empirically. Experiments have shown that under different simulation configurations, setting $\eta_{th}=0.2$ could achieve better trade-off between convergence speed and robustness against error propagation. Line 12 represents the EM update in the outer iteration. In addition, Line 13-14 represent the message initialization for the next round of inner iterations after the $\tau$-th EM update. Finally, the BR-MP-EM algorithm outputs the UAD result $\hat{\mathbf{a}}=\left\{\hat{\alpha}_k,\forall k\right\}$ and CE result $\hat{\mathbf{h}}=\left\{\hat{\alpha}_k\mu^\text{dec}_{k}(N_{in}),\forall k\right\}$.
\begin{remark}\label{generality}
	The proposed BR-MP-EM algorithm treats the random phase shift $\phi_\delta$ and propagation fading $h_r$ as hyper-parameters, and updates these hyper-parameters with the EM method. It is worth mentioning that the EM method does not rely on specific distributions of $\phi_\delta$ and $h_r$. Instead, only the average measured propagation fading $\overline{h}_r$ is needed for this algorithm.   
\end{remark}
\begin{remark}\label{hyperaffect}
	In (\ref{channelmodel}) we assume the channel impairments $h_re^{j\phi_{\delta}}$ affect both the scattering component and the LoS component. However, this assumption does not sacrifice any loss of generality for the BR-MP-EM algorithm. That is, if the channel impairments only affect the LoS component, only some minor modifications are required on the BR-MP-EM algorithm in Algorithm \ref{alg:BRMPEM}. The details are provided in Appendix \ref{EMmodif}. 
\end{remark}
\begin{remark}\label{largearea}
It is assumed in Algorithm \ref{alg:BRMPEM} that $K$ devices experience the same channel impairment $h_re^{j\phi_{\delta}}$. However, the BR-MP-EM algorithm can be readily generalized to the case when devices experience different channel impairments. Assume that $K$ devices are divided into $G$ groups by geographical proximity, and only the devices in the same group experience the same channel impairment. Since IoT devices are assumed stationary, the group membership can be acquired by the satellite. Then, in order to generalize the BR-MP-EM algorithm, we only need to independently perform the EM update (Line 12 of Algorithm \ref{alg:BRMPEM}) for each group of devices.
\end{remark}
\begin{table}
	\caption{Computational Complexity in Each Iteration}\vspace{0.2cm}
	\centering
	\begin{tabular}{|c|c|c|c|}
		\hline
		Message&Eqn.&Real Multi. / Div.&Exp. / Log.\\\hline
		$\mu^*_{lk}(t)$&(\ref{equimuvar})&$6KL$&0\\\hline
		$v^*_{lk}(t)$&(\ref{equimuvar})&$5KL$&0\\\hline
		$\mu^\text{sn}_{l\to k}(t)$&(\ref{SNupdateGaussian})&$4KL$&0\\\hline
		$v^\text{sn}_{l\to k}(t)$&(\ref{SNupdateGaussian})&$KL$&0\\\hline
		$\mu^1_{lk}(t)$&(\ref{SNupdateBernoulli})&$4KL$&0\\\hline
		$v^1_{lk}(t)$&(\ref{SNupdateBernoulli})&$KL$&0\\\hline
		$l^\text{sn}_{l\to k}(t)$&(\ref{LLR})&$7KL$&$KL$\\\hline
		$v^\text{vn}_{k\to l}(t+1)$&(\ref{VNupdateGaussian})&$2KL+K$&0\\\hline
		$\mu^\text{vn}_{k\to l}(t+1)$&(\ref{VNupdateGaussian})&$4KL+2K$&0\\\hline
		$p^\text{vn}_{k\to l}(t+1)$&(\ref{reverseLLRtrans})&$KL$&$KL$\\\hline
		$v^\text{dec}_{k}(t)$&(\ref{VNCE})&$K$&0\\\hline
		$\mu^\text{dec}_{k}(t)$&(\ref{VNCE})&$2K$&0\\\hline
		$l^\text{ce}_k(t)$&(\ref{ceforllr})&$7K$&$K$\\\hline
		$M(\tau)$&(\ref{solution})&$7\|\mathcal{K^+}\|$&0\\\hline
		$N(\tau)$&(\ref{solution})&$3\|\mathcal{K^+}\|$&0\\\hline
	\end{tabular}
	\label{comple}
\end{table}
\subsection{Computational Complexity}
One remarkable property of the message passing algorithms is that the overall processing can be decomposed into many local computations, which are executed in parallel at different SNs and VNs on the factor graph. Therefore, the proposed BR-MP-EM algorithm exhibits low computational complexity, which is favorable for reducing the processing delay at the satellite receiver. The computational complexity of the proposed BR-MP-EM algorithm is analyzed as follows.

We evaluate the computational complexity by the number of real-number multiplication (or division) operations and the number of exponential (or logarithm) operations. In each inner iteration and outer iteration, the complexity for calculating related messages is listed in Table. \ref{comple}. It is shown that the computational complexity is mainly incurred by the Bernoulli-Rician message passing in the inner iterations, while the operations in the EM update are almost negligible. According to Table \ref{comple}, each inner iteration costs about $35KL$ multiplications (or divisions) and $2KL$ exponential (or logarithm) operations. Furthermore, as in \cite{GQH}, if we approximate the extrinsic messages $v^\text{vn}_{k\to l}(t+1)$, $\mu^\text{vn}_{k\to l}(t+1)$, and $l^\text{vn}_{k\to l}(t+1)$ in each VN update with the following full messages $v^\text{full}_{k}(t+1)$, $\mu^\text{full}_{k}(t+1)$, and $l^\text{full}_{k}(t+1)$,
\begin{equation}\label{approx}
\begin{split}
v^\text{full}_{k}(t+1)\overset{\Delta}{=}&\left[\left(v^\text{pri}_k(\tau)\right)^{-1}+\sum_{i\in\mathcal{L}}\left(v^\text{sn}_{i\to k}(t)\right)^{-1}\right]^{-1},\\
\mu^\text{full}_{k}(t+1)\overset{\Delta}{=}&v^\text{full}_{k}(t+1)\left[ \frac{\mu^\text{pri}_k(\tau)}{v^\text{pri}_k(\tau)} + \sum_{i\in\mathcal{L}}\frac{\mu^\text{sn}_{i\to k}(t)}{v^\text{sn}_{i\to k}(t)} \right],\\
l^\text{full}_{k}(t+1)\overset{\Delta}{=}&l_0+\sum_{l\in\mathcal{L}}l^\text{sn}_{l\to k}(t),
\end{split}
\end{equation}
then the computational complexity of each inner iteration will be reduced to $31KL$ multiplications and $KL$ exponential operations.
Therefore, the overall complexity of the BR-MP-EM algorithm is as low as $\mathcal{O}(KLN_{it})$ multiplications and $\mathcal{O}(KLN_{it})$ exponential operations, where $N_{it}=N_{in}N_{out}$ is the total number of iterations. 
\section{Simulation Results}\label{capability}
Without loss of any generality, for the following simulations, we assume that the propagation fading $h_r$ follows a log-normal distribution \cite{RLN} with pdf
\begin{equation}\label{lognormal}
f_r(h_r|\mu_r,\sigma_r)\!=\!\frac{1}{\sqrt{2\pi}C\sigma_rh_r}\exp\left[-\frac{1}{2}\left(\frac{\ln h_r\!-\!\mu_r}{C\sigma_r}\right)^2\right]
\end{equation}
where $C\overset{\Delta}{=}{(\ln 10)}/{20}$ \cite{RLN}, and the average measured propagation fading is set as $\overline{h}_r=\mathbf{E}[h_r]=e^{\mu_r+(C\sigma_r)^2/2}$. In addition, the phase shift $\phi_\delta$ is randomly drawn from a Gaussian distribution $\mathcal{N}(0,\sigma_\delta^2)$. As pointed out in Remark \ref{generality}, the EM update does not rely on specific distributions of $h_r$ and $\phi_\delta$. Therefore, above-mentioned assumptions are merely for the convenience of simulations. In the following simulations, we investigate the UAD and CE performances of our proposed BR-MP-EM algorithm under different configurations of the terrestrial-satellite GF-RA system.
\begin{figure*}
	\centering
	\includegraphics[width=0.94\textwidth]{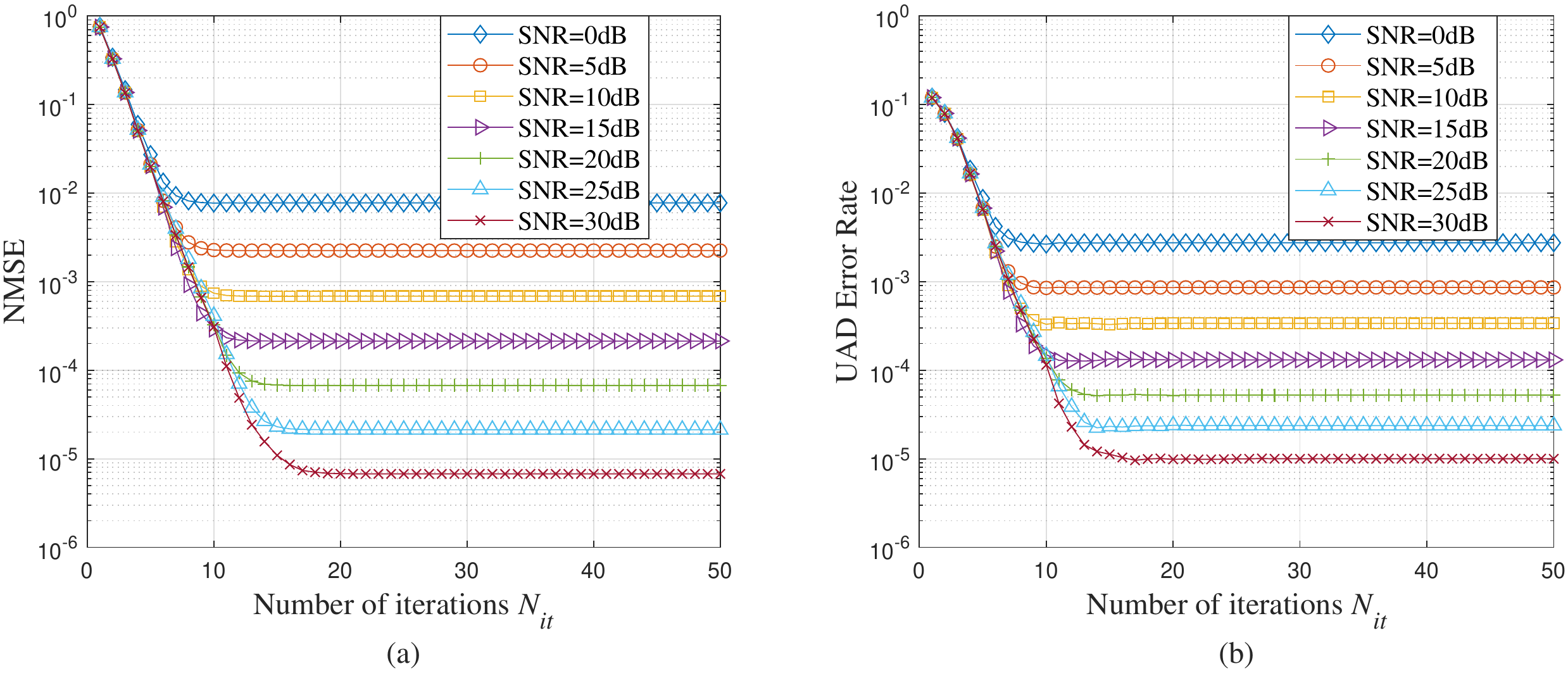}\vspace{-0.3cm}
	\caption{UAD and CE performances of the BR-MP-EM algorithm under various SNR. Related parameters are ($K=500, L=200, p_a=0.1, N_{in}=10, N_{out}=5, \eta_{th}=0.2, \phi^\text{los}_k\sim\mathcal{U}[-\pi, \pi],  |h^\text{los}_k|^2\sim\mathcal{U}[0.6, 0.7], v^\text{ray}_k\sim\mathcal{U}[0.2, 0.25], \sigma_\delta=\frac{\pi}{8}, \sigma_r=1.0, \mu_r=0.13$ \cite{Loo}).\vspace{-0.3cm}}\label{changeSNR}
\end{figure*}
\begin{figure*}
	\centering
	\includegraphics[width=0.94\textwidth]{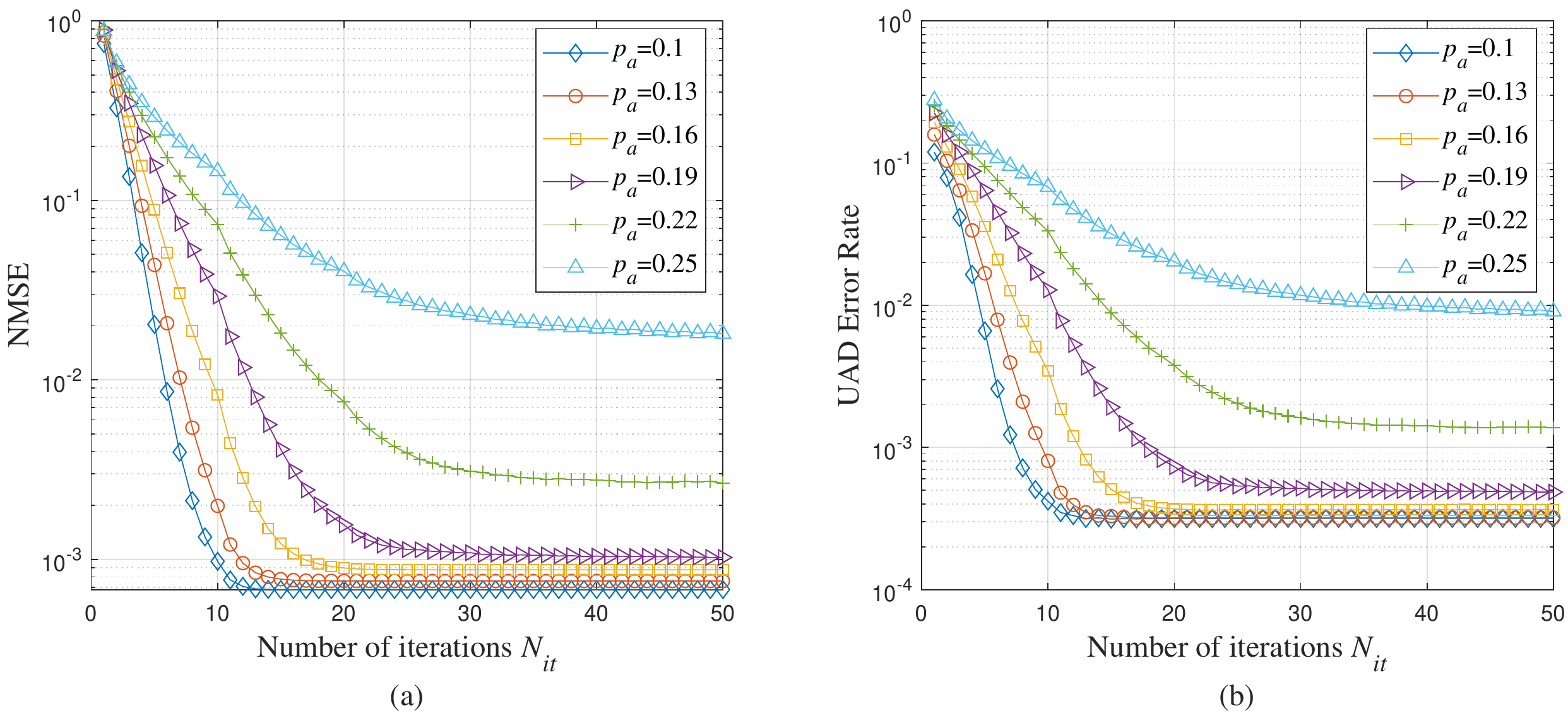}\vspace{-0.3cm}
	\caption{UAD and CE performances of the BR-MP-EM algorithm with various activation probabilities $p_a$. Related parameters are ($K=500, L=200, \text{SNR}=10\text{dB}, N_{in}=10, N_{out}=5, \eta_{th}=0.2, \phi^\text{los}_k\sim\mathcal{U}[-\pi, \pi], |h^\text{los}_k|^2\sim\mathcal{U}[0.6, 0.7], v^\text{ray}_k\sim\mathcal{U}[0.2, 0.25], \sigma_\delta=\frac{\pi}{8}, \sigma_r=1.0, \mu_r=0.13$ \cite{Loo}).\vspace{-0.3cm}}\label{changepa}
\end{figure*}
\subsection{Impacts of Signal-to-Noise Ratio}
Firstly, we investigate the performances of our proposed BR-MP-EM algorithm under various signal-to-noise ratio (SNR), which is defined as $\text{SNR}\overset{\Delta}{=}\log_{10}\frac{1}{\sigma_n^2}$ and $\sigma_n^2$ represents the noise variance of $n_l$ in (\ref{equinoise}). The numerical simulation results for the CE and UAD performances of the BR-MP-EM algorithm are illustrated in Fig. \ref{changeSNR}(a) and Fig. \ref{changeSNR}(b), respectively. In addition, we employ the normalized mean square error (NMSE) $\|\hat{\mathbf{h}}\odot\hat{\mathbf{a}}-\mathbf{h}\odot\mathbf{a}\|^2/\|\mathbf{h}\odot\mathbf{a}\|^2$ to evaluate the CE accuracy.

It is shown in Fig. \ref{changeSNR} that our proposed BR-MP-EM algorithm works within a wide range of SNR. In addition, both the CE and UAD accuracy of the BR-MP-EM algorithm gets improved with the increase of the SNR. The performance improvement exhibits a linear relationship with SNR in the logarithmic scale. Furthermore, under the system configurations considered in Fig. \ref{changeSNR}, the proposed BR-MP-EM algorithm can always guarantee convergence with less than 20 iterations, indicating low processing delay at the satellite receiver.  
\begin{figure*}
	\centering
	\includegraphics[width=0.94\textwidth]{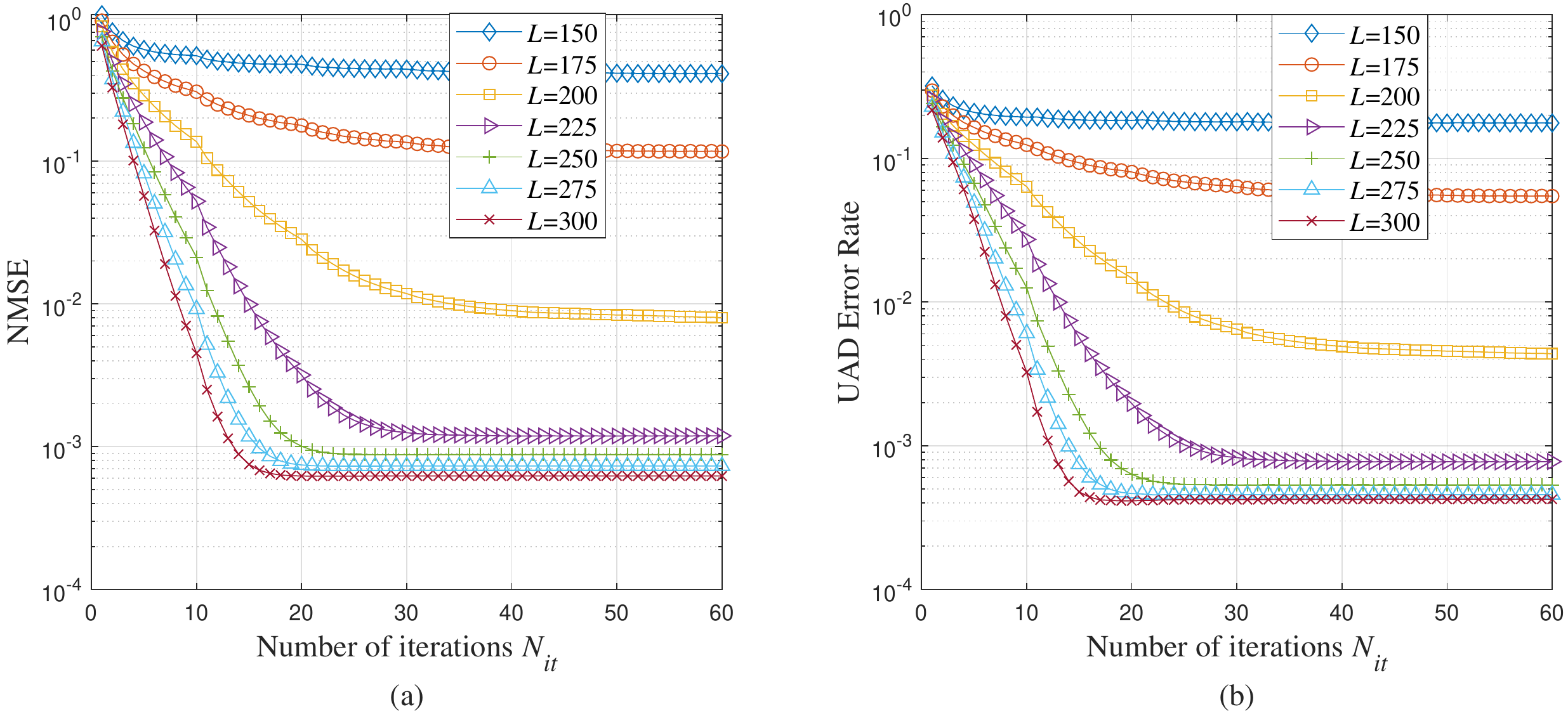}\vspace{-0.3cm}
	\caption{UAD and CE performances of the BR-MP-EM algorithm with various pilot length $L$. Related parameters are ($K=500, p_a=0.25, \text{SNR}=10\text{dB}, N_{in}=10, N_{out}=6, \eta_{th}=0.2, \phi^\text{los}_k\sim\mathcal{U}[-\pi, \pi], |h^\text{los}_k|^2\sim\mathcal{U}[0.6, 0.7], v^\text{ray}_k\sim\mathcal{U}[0.2, 0.25], \sigma_\delta=\frac{\pi}{8}, \sigma_r=1.0, \mu_r=0.13$ \cite{Loo}).\vspace{-0.3cm}}\label{changeL}
\end{figure*}
\begin{figure*}
	\centering
	\includegraphics[width=0.94\textwidth]{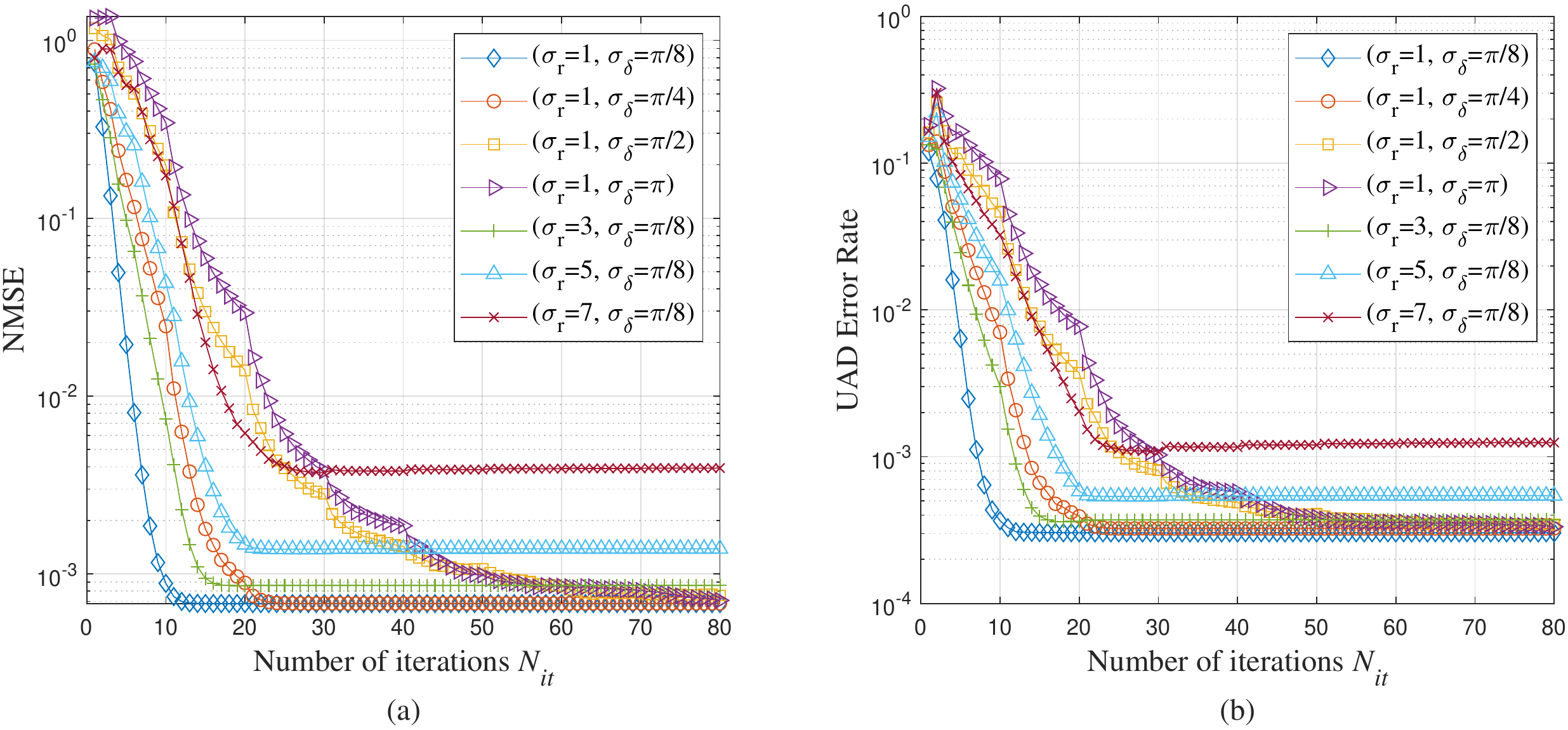}\vspace{-0.3cm}
	\caption{UAD and CE performances of the BR-MP-EM algorithm with various hyper-parameter variance $\sigma^2_r$ and $\sigma^2_\delta$. Related parameters are ($K=500, L=200, p_a=0.1, \text{SNR}=10\text{dB}, N_{in}=10, N_{out}=8, \eta_{th}=0.2, \phi^\text{los}_k\sim\mathcal{U}[-\pi, \pi], |h^\text{los}_k|^2\sim\mathcal{U}[0.6, 0.7], v^\text{ray}_k\sim\mathcal{U}[0.2, 0.25], \mu_r=0.13$ \cite{Loo}).\vspace{-0.3cm}}\label{changeHyper}
\end{figure*}
\subsection{Impacts of Activation Probability}
In order to show the applicability of the BR-MP-EM algorithm in different IoT applications, we investigate the UAD and CE accuracy of our proposed BR-MP-EM algorithm with various activation probabilities $p_a$. Related simulation results are illustrated in Fig. \ref{changepa}. 

It is shown in Fig. \ref{changepa} that the UAD and CE performances exhibit similar characteristics with the increase of $p_a$. Firstly, when $p_a$ increases from 0.1 to 0.19, the final UAD and CE performances do not degrade seriously. However, with a larger $p_a$, more iterations are required for the BR-MP-EM algorithm to converge. When $p_a$ further increases to 0.22 and 0.25, the UAD and CE performances are undermined since this GF-RA system is overwhelmed with more active devices. As shown by the following simulations, this RA congestion problem can be potentially solved by employing longer pilot sequences, i.e., larger $L$.      
\subsection{Impacts of Pilot Length}
As pointed out in Remark \ref{pilot}, the pilot length $L$ is related to the access efficiency. Therefore, given the desired CE and UAD accuracy, it is preferred to use pilot sequences that are as short as possible. The performances of the BR-MP-EM algorithm are then investigated with various pilot length $L$, and the simulation results are illustrated in Fig. \ref{changeL}.

Similar to the observations in Fig. \ref{changepa}, it is shown in Fig. \ref{changeL} that there is a critical value of pilot length, i.e. $L=225$. When $L$ is smaller than this critical value, the joint UAD and CE performances deteriorate severely. On the other hand, when $L$ is larger than this critical value, the convergence speed of the BR-MP-EM algorithm is improved. But the UAD and CE accuracy cannot be significantly improved by further increasing $L$, since the performance of the BR-MP-EM algorithm is also limited by other parameters such as SNR.

\subsection{Robustness Against Channel Impairments}
As mentioned above, the proposed BR-MP-EM algorithm employs EM update to estimate unknown hyper-parameters $h_r$ and $\phi_\delta$ that are caused by channel impairments. Then, we investigate the robustness of the proposed BR-MP-EM algorithm when the hyper-parameters $h_r$ and $\phi_\delta$ have different variances, i.e. $\sigma^2_r$ and $\sigma^2_\delta$. To evaluate the performance of the EM update, we additionally define the estimation mean square error $\Delta_\text{hp}=|h_re^{j\phi_\delta}-\hat{h}_re^{j\hat{\phi}_\delta}|^2$ to measure the accuracy of the estimated hyper-parameters $\hat{h}_r$ and $\hat{\phi}_\delta$. The simulation results are shown in Fig. \ref{changeHyper}. 

It is shown in Fig. \ref{changeHyper} that when $\sigma_r=1$ and $\sigma_\delta$ increases from $\pi/8$ to $\pi$, more iterations are required by the BR-MP-EM algorithm to achieve convergence. However, the proposed BR-MP-EM algorithm could always guarantee the same UAD and CE accuracy after convergence, even when $\sigma_\delta=\pi$. In addition, $\Delta_\text{hp}$ could always reach $7.5\times10^{-3}$ for various $\sigma_\delta$ when $\sigma_r=1$. On the other hand, if we fix $\sigma_\delta=\pi/8$ and increase $\sigma_r$ from 1 to 7, the TSL suffers from greater uncertainty of the propagation fading $h_r$. It is shown that the UAD and CE accuracy will gradually degrade. In addition, the variance of the channel impairment, i.e.  $\mathbf{Var}\left[h_re^{j\phi_{\delta}}\right]$ increases from $2.05\times10^{-1}$ to $2.64$ when $\sigma_r$ increases from 1 to 7. However, $\Delta_\text{hp}$ of our proposed algorithm only increases from $7.5\times10^{-3}$ to $2.58\times10^{-2}$, which proves the robustness of our proposed BR-MP-EM algorithm against unknown channel impairments.
\begin{figure}
	\centering
	\includegraphics[width=0.45\textwidth]{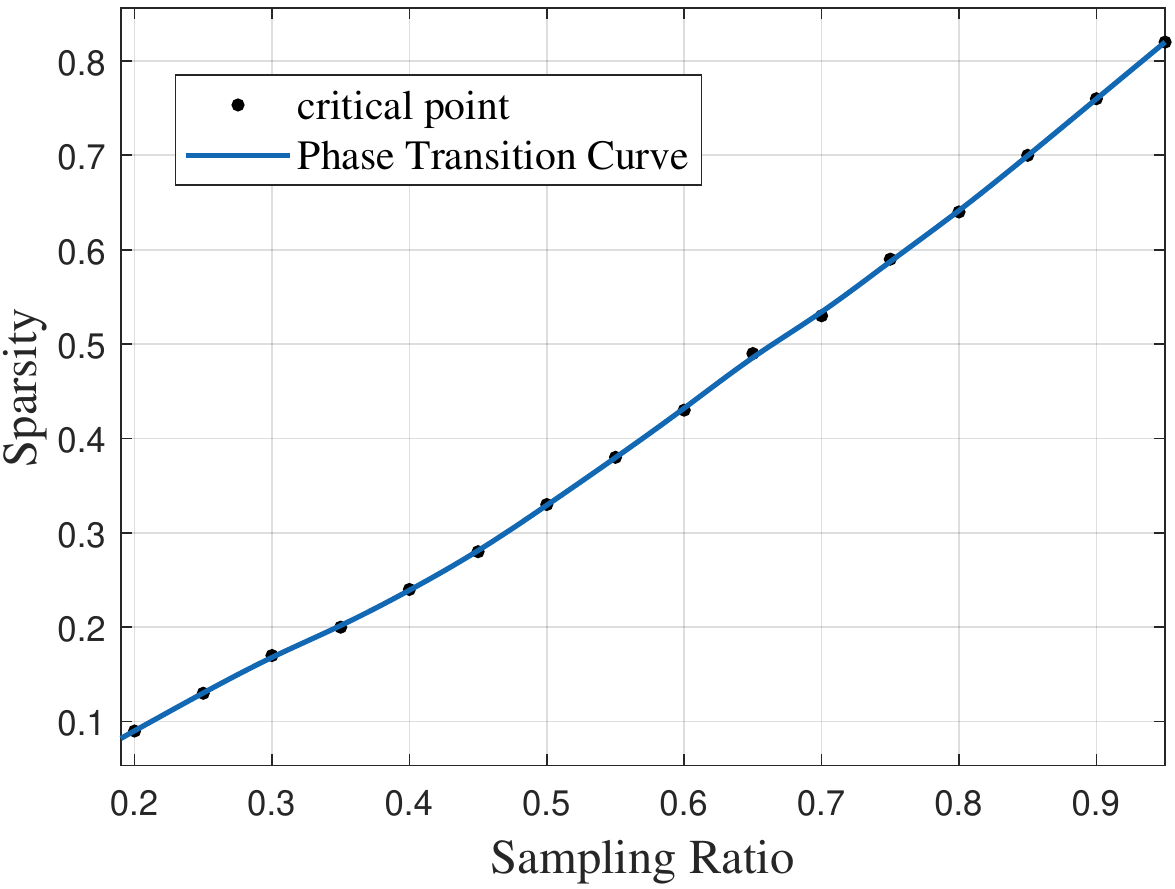}
	\caption{Phase transition curve for the proposed BR-MP-EM algorithm. Related parameters are ($K=500, \text{SNR}=40\text{dB}, N_{in}=10, N_{out}=15, \eta_{th}=0.2, \phi^\text{los}_k\sim\mathcal{U}[-\pi, \pi], |h^\text{los}_k|^2\sim\mathcal{U}[0.6, 0.7], v^\text{ray}_k\sim\mathcal{U}[0.2, 0.25], \sigma_\delta=\frac{\pi}{8}, \sigma_r=1.0, \mu_r=0.13$ \cite{Loo}).}\label{PTC}
\end{figure}
\subsection{Phase Transition for Sparse Recovery}
The joint UAD and CE problem (\ref{problemformulation}) is equivalent to a recovery problem for a sparse vector $\mathbf{h}_{K\times 1}\odot\mathbf{a}_{K\times 1}$. In addition, it is shown in Fig. \ref{changepa} and Fig. \ref{changeL} that there is a particular threshold for $p_a$ and $L$ so that the proposed BR-MP-EM algorithm can generate accurate UAD and CE result. For example, if $p_a\leq0.19$ in Fig. \ref{changepa} or $L\ge225$ in Fig. \ref{changeL}, the UAD and CE accuracy  can be well guaranteed. Otherwise, the performances of the proposed BR-MP-EM algorithm deteriorate rapidly. Define $p_a$ as the \emph{sparsity} of the vector $\mathbf{h}_{K\times 1}\odot\mathbf{a}_{K\times 1}$, and $L/K$ as the \emph{sampling ratio}. We then investigate the phase transition for the sparse recovery problem (\ref{problemformulation}), i.e., we find the threshold of $p_a$ and $L$ that  guarantees the UAD and CE accuracy.

It is shown from Fig. \ref{changeSNR} to Fig. \ref{changeHyper} that the UAD and CE performance curves exhibit similar characteristics under different system configurations. Therefore, we focus on the UAD accuracy, and further define \emph{successful recovery} as the case when the average UAD error rate is smaller than $1/K$. Then, via numerical simulations, we find the critical grid points ($L/K,p_a$) so that the BR-MP-EM algorithm can guarantee \emph{successful recovery}. The simulation results are illustrated in Fig. \ref{PTC}, and the phase transition curve (PTC) is obtained by curve-fitting on these critical grid points. The grid points below the PTC indicate the case when successful recovery can be guaranteed. It is shown that on the PTC, the sampling ratio grows almost linearly with the sparsity. That is, with longer pilot sequences, the proposed BR-MP-EM algorithm can support more simultaneously activated devices.
\begin{figure}
	\centering
	\includegraphics[width=0.45\textwidth]{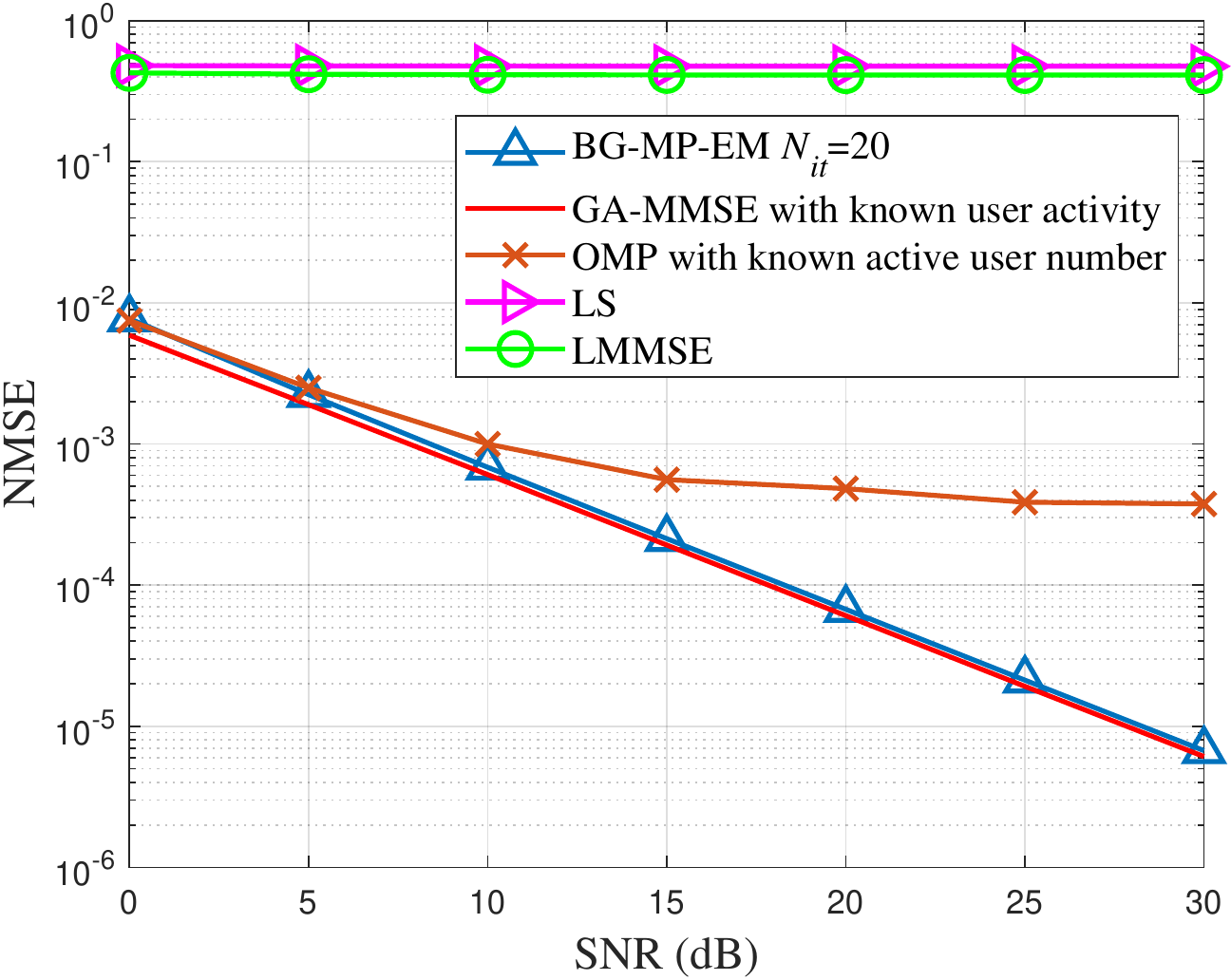}
	\caption{NMSE performance comparison between the proposed BR-MP-EM algorithm and LS estimator, LMMSE estimator, OMP estimator (with known number of active devices), and GA-MMSE estimator (with known device activity and hyper-parameters). The NMSE performance of the GA-MMSE estimator serves as the performance lower bound. Related parameters are ($K=500, L=200, p_a=0.1, N_{in}=10, N_{out}=2, \eta_{th}=0.2, \phi^\text{los}_k\sim\mathcal{U}[-\pi, \pi], |h^\text{los}_k|^2\sim\mathcal{U}[0.6, 0.7], v^\text{ray}_k\sim\mathcal{U}[0.2, 0.25], \sigma_\delta=\frac{\pi}{8}, \sigma_r=1.0, \mu_r=0.13$ \cite{Loo}).}\label{Comparison}
\end{figure}
\subsection{Performance Comparison}
Finally, we present the NMSE performance comparison between our proposed BR-MP-EM algorithm and other methods including the linear minimum square error (LMMSE) estimator, the least square (LS) estimator, and the orthogonal matching pursuit (OMP) estimator (with known active device number). Furthermore, we also present the NMSE performance of the genie-aided MMSE (GA-MMSE) estimator, which is assumed with known device activity and hyper-parameters $h_r$ and $\phi_\delta$. The NMSE performance of the GA-MMSE estimator serves as the lower-bound \cite{BGMP}. The comparison results are illustrated in Fig. \ref{Comparison}.

It is shown in Fig. \ref{Comparison} that with only $N_{it}=20$ iterations, the NMSE performance of our proposed BR-MP-EM algorithm can always closely approach the lower bound in a wide range of SNR. In contrast, the OMP estimator diverges from the lower bound with the increase of SNR, i.e., increasing SNR will not lead to lower NMSE for the OMP estimator. This divergence of the OMP estimator can be explained by the non-orthogonality of the Gaussian pilot sequences, especially when the pilot length $L$ is much smaller than the device number $K$. In addition, it is shown that the LMMSE estimator and the LS estimator fail to work when they are directly applied to address the joint UAD and CE problem in this terrestrial-satellite GF-RA system. 
\section{Conclusions}\label{conclusions}
A terrestrial-satellite GF-RA system was considered in this paper to support IoT applications with LEO satellites. A BR-MP-EM algorithm was proposed to address the joint UAD and CE problem in this GF-RA system. This BR-MP-EM algorithm is divided into inner iterations and outer iterations. In the inner iterations, the Bernoulli-Rician message passing was derived for the joint UAD and CE problem. In the outer iterations, the EM update was derived to estimate the channel impairment-related hyper-parameters, i.e., the propagation fading and the phase shift. Finally, simulation results were presented to show the joint UAD and CE accuracy, as well as the robustness of the proposed BR-MP-EM algorithm against unknown channel impairments.  
\begin{appendices}
\section{Derivation of EM Update for Hyper-Parameters}\label{derivation}
According to (\ref{priGaussianpdf}), we have 
\begin{equation}\label{lnexpansion}
\begin{split}
\ln f^\text{pri}(h_k;h_r,\phi_{\delta})=&-\ln({\pi v^\text{ray}_k})+\frac{2{\textbf{R}}[h_kh^\text{los}_ke^{-j(\phi^\text{los}_k+\phi_\delta)}]}{h_rv^\text{ray}_k}\\
&-\frac{|h_k|^2+h^2_r(h^\text{los}_k)^2}{h^2_rv^\text{ray}_k}-2\ln{h_r}
\end{split}
\end{equation}
where $\textbf{R}[\cdot]$ represents the real-part of a complex number.
Differentiate (\ref{lnexpansion}) with respect to $h_r$, we have
\begin{equation}\label{partialhr}
\frac{\partial\ln f^\text{pri}(h_k;h_r,\phi_{\delta})}{\partial h_r}\!\!=\!-\!\frac{2}{h_r}\!+\!\frac{2|h_k|^2}{v^\text{ray}_kh_r^3}\!-\!\frac{2{\textbf{R}[h_kh^\text{los}_ke^{-j(\phi^\text{los}_k\!+\phi_\delta)}]}}{v^\text{ray}_kh_r^2}
\end{equation}
Multiplying ${-h_r^3}/{2}$ on both sides will not change the equality of the first equation in (\ref{equationset}). Therefore, we have
\begin{equation}\label{hrchange}
\begin{split}
g^h_k\overset{\Delta}{=}&\frac{-h_r^3}{2}\frac{\partial\ln f^\text{pri}(h_k;h_r,\phi_{\delta})}{\partial h_r}\\
=&h_r^2-h_r\frac{{\textbf{R}[h_kh^\text{los}_ke^{-j(\phi^\text{los}_k\!+\phi_\delta)}]}}{v^\text{ray}_k}+\frac{|h_k|^2}{v^\text{ray}_k}
\end{split}
\end{equation}
where $g^h_k$ is defined for notational simplicity. Substitute (\ref{partialhr}) and (\ref{hrchange}) into the first equation of (\ref{equationset}), we have
\begin{equation}\label{equationh}
\begin{split}
&\!\!\int\ldots\int{\prod\limits_{k\in\mathcal{K}^+}f^\text{pos}_\tau(h_k)\left(\sum\limits_{k\in\mathcal{K}^+}g^h_k \right)}\prod\limits_{k\in\mathcal{K}^+}dh_k\\
\overset{(d)}{=}&\!\!\!\sum\limits_{k\in\mathcal{K}^+}\left(\int{f^\text{pos}_\tau(h_k)g^h_k}dh_k\right)\\
\overset{(e)}{\!=\!}&\!\!\!\sum\limits_{k\in\mathcal{K}^+}\!\!\left(\!\!h_r^2\!-\!h_r\frac{{\textbf{R}[\mu^\text{dec}_k(t)h^\text{los}_ke^{-j(\phi^\text{los}_k\!+\phi_\delta)}]}}{v^\text{ray}_k}\!+\!\!\frac{|\mu^\text{dec}_k(t)|^2\!+\!\!v^\text{dec}_k(t)}{v^\text{ray}_k}\right)\\
=&0
\end{split}
\end{equation}
where equation ($d$) of (\ref{equationh}) is obtained by the independence among different device $k\in\mathcal{K}^+$, and equation ($e$) of (\ref{equationh}) is obtained by taking the expectation of $g_k^h$ over $h_k$ with posterior Gaussian pdf $f^\text{pos}_\tau(h_k)$.

Differentiate (\ref{lnexpansion}) with respect to $\phi_\delta$, we have
\begin{equation}\label{partialphi}
\frac{\partial\ln f^\text{pri}(h_k;h_r,\phi_{\delta})}{\partial \phi_\delta}=\frac{{2\textbf{R}[h_kh^\text{los}_ke^{-j(\phi^\text{los}_k\!+\phi_\delta+\frac{\pi}{2})}]}}{v^\text{ray}_kh_r}
\end{equation}
Similarly, we define $g^\phi_k\overset{\Delta}{=}\frac{h_r}{2}\frac{\partial\ln f^\text{pri}(h_k;h_r,\phi_{\delta})}{\partial \phi_\delta}$ for notational convenience. Substitute $g^\phi_k$ and (\ref{partialphi}) into the second equation of (\ref{equationset}), we have
\begin{equation}\label{equationphi}
\begin{split}
&\int\ldots\int{\prod\limits_{k\in\mathcal{K}^+}f^\text{pos}_\tau(h_k)\left(\sum\limits_{k\in\mathcal{K}^+}g^\phi_k\right)}\prod\limits_{k\in\mathcal{K}^+}dh_k\\
=&\sum\limits_{k\in\mathcal{K}^+}\left(\int{f^\text{pos}_\tau(h_k)g^\phi_k}dh_k\right)\\
=&\sum\limits_{k\in\mathcal{K}^+}\left(\frac{\textbf{R}[\mu^\text{dec}_k(t)h^\text{los}_ke^{-j(\phi^\text{los}_k\!+\phi_\delta+\frac{\pi}{2})}]}{v^\text{ray}_k}\right)=0
\end{split}
\end{equation}
After some simple mathematical manipulations over (\ref{equationh}) and (\ref{equationphi}), we can obtain $\hat{\phi}_\delta(\tau+1)$ and $\hat{h}_r(\tau+1)$ by solving the following equation set
\begin{equation}\label{finalequationset}
\left\{ \begin{split}
\!h_r^2\!+\!h_r\!\!\left\langle\!\frac{\textbf{R}[\mu^\text{dec}_k\!(t)h^\text{los}_ke^{-j(\phi^\text{los}_k\!+\!\phi_\delta)}]}{v^\text{ray}_k}\!\right\rangle\!\!-\!\!\left\langle\!\frac{|\mu^\text{dec}_k\!(t)|^2\!\!+\!\!v^\text{dec}_k\!(t)}{v^\text{ray}_k}\!\right\rangle\!\!=\!\!0\\
\!\left\langle\!\!\frac{\mu^\text{dec}_k\!(t)h^\text{los}_ke^{-j\phi^\text{los}_k}}{v^\text{ray}_k}\!\!\right\rangle e^{-j\phi_\delta}\!=\!\left\langle\frac{[\mu^\text{dec}_k(t)]^Ch^\text{los}_ke^{j\phi^\text{los}_k}}{v^\text{ray}_k}\right\rangle e^{j\phi_\delta}.  
\end{split}
\right.
\end{equation} 
where $[\mu^\text{dec}_k(t)]^C$ represents the complex conjugate of ${\mu^\text{dec}_k(t)}$. It is noted that the mathematical solution to (\ref{solution}) is not unique, and we can actually obtain two sets of solutions, as in  (\ref{solution1}) and  (\ref{solution2})
\begin{equation}\label{solution1}
\left\{ \begin{split}
\hat{\phi}_\delta&=\angle M\\
\hat{h}_r&=\frac{-|M|+\sqrt{|M|^2+4N}}{2} 
\end{split}
\right.
\end{equation} 
\begin{equation}\label{solution2}
\left\{ \begin{split}
\hat{\phi}_\delta&=\angle M+\pi\ \textbf{or} \ \angle M-\pi\\
\hat{h}_r&=\frac{|M|+\sqrt{|M|^2+4N}}{2} 
\end{split}
\right.
\end{equation} 
where $M=\left\langle\left(\mu^\text{dec}_{k}(t)h^\text{los}_ke^{-j\phi^\text{los}_k}\right)/{v^\text{ray}_k}\right\rangle$, $N=\left\langle\left(v^\text{dec}_{k}(t)+|\mu^\text{dec}_{k}(t)|^2\right)/{v^\text{ray}_k}\right\rangle$. However, it is emphasized that we can rule out the solution (\ref{solution2}) by looking into the physical interpretation of $\angle M$. We rewrite $M$ as
\begin{equation}\label{M}
\begin{split}
M&=\frac{1}{\|\mathcal{K^+}\|}\sum_{k\in\mathcal{K^+}}\frac{h^\text{los}_k}{v^\text{ray}_k}\mu^\text{dec}_{k}(t)e^{-j\phi^\text{los}_k}\\
&=\frac{1}{\|\mathcal{K^+}\|}\sum_{k\in\mathcal{K^+}}\frac{h^\text{los}_k|\mu^\text{dec}_{k}(t)|}{v^\text{ray}_k}e^{j(\phi^\text{dec}_k(t)-\phi^\text{los}_k)}\\
&\overset{(f)}{=}\frac{1}{\|\mathcal{K^+}\|}\sum_{k\in\mathcal{K^+}}\frac{h^\text{los}_k|\mu^\text{dec}_{k}(t)|}{v^\text{ray}_k}m_k
\end{split}
\end{equation}
where $\phi^\text{dec}_k(t)\overset{\Delta}{=}\angle \mu^\text{dec}_{k}(t)$. As shown in  equation ($f$) of (\ref{M}), $M$ can be interpreted as a weighted mean-value of $m_k\overset{\Delta}{=}e^{j(\phi^\text{dec}_k(t)-\phi^\text{los}_k)}$ over all the devices $k\in\mathcal{K^+}$, and the term ${h^\text{los}_k|\mu^\text{dec}_{k}(t)|}/{v^\text{ray}_k}$ serves as the weight for $m_k$. Now we consider the estimate $\mu^\text{dec}_{k}(t)$ of $h_k$, which is passed from the left sub-graph. Assume that the estimate $\mu^\text{dec}_{k}(t)$ is ideally equal to $h_k$, then  $\mathbf{E}\left[\phi^\text{dec}_k(t)\right]=\phi^\text{los}_k+\phi_\delta$ according to $(\ref{channelmodel})$. Therefore, $\mathbf{E}\left[\angle M\right]=\mathbf{E}\left[\angle m_k\right]=\phi_\delta$ according to equation ($f$) of (\ref{M}). In this way, the solution (\ref{solution1}) can be interpreted, and we can rule out the other solution (\ref{solution2}). 
\section{Modifications When Channel Impairments  Only Affect LoS Component}\label{EMmodif}
If the channel impairments affect only the LoS component, the terrestrial-satellite channel for device $k$ is changed from (\ref{channelmodel}) to
\begin{equation}\label{modichannelmodel}
h_k=h^\text{ray}_ke^{j\phi^\text{ray}_k}+h_re^{j\phi_{\delta}}h^\text{los}_ke^{j\phi^\text{los}_k}
\end{equation}
Compared with the BR-MP-EM algorithm described in Algorithm \ref{alg:BRMPEM}, only some minor modifications are required if the channel model (\ref{modichannelmodel}) is considered instead of (\ref{channelmodel}). Related modifications are explained as follows.

If the channel impairments only affect the LoS component, then the scattering component in $h_k$ has constant variance $v^\text{ray}_k$. Therefore, for the Bernoulli-Rician message passing in the inner iterations, the variable ${v^\text{pri}_k(\tau)}$ is replaced with a constant $v^\text{ray}_k$ in (\ref{VNupdateGaussian}), (\ref{VNCE}), and (\ref{ceforllr}). For the EM update in the outer iterations, the function $f^\text{pri}(h_k;h_r,\phi_{\delta})$ in (\ref{priGaussianpdf}) is now modified as $f^\text{pri}_\text{modi}(h_k;h_r,\phi_{\delta})$
\begin{equation}\label{modipriGaussianpdf}
f^\text{pri}_\text{modi}(h_k;h_r,\phi_{\delta})\overset{\Delta}{=}f(h_k|h_rh^\text{los}_ke^{j(\phi^\text{los}_k+\phi_{\delta})}, v^\text{ray}_k).
\end{equation}
Then we perform this replacement in (\ref{EMmoreuser}) and (\ref{equationset}). According to the derivations in Appendix \ref{derivation}, we have the following modified solution to the EM update
\begin{equation}\label{modisolution}
\left\{ \begin{split}
\hat{\phi}^\text{modi}_\delta&=\angle\left\langle\frac{\mu^\text{dec}_{k}(t)h^\text{los}_ke^{-j\phi^\text{los}_k}}{v^\text{ray}_k}\right\rangle\\
\hat{h}^\text{modi}_r&=\frac{\left|\left\langle\mu^\text{dec}_{k}(t)h^\text{los}_ke^{-j\phi^\text{los}_k}/v^\text{ray}_k\right\rangle\right|}{\left\langle\left(h^\text{los}_k\right)^2/v^\text{ray}_k\right\rangle}
\end{split}
\right.
\end{equation} 
\end{appendices}
\ifCLASSOPTIONcaptionsoff
  \newpage
\fi


%\bibliographystyle{IEEEtran}

\begin{thebibliography}{1}
\bibitem{IoT1}
L. Atzori, A. Iera, and G. Morabito, ``The Internet of Things: A survey,'' \emph{Comput. Netw.}, vol. 54, no. 15, pp.2787-2805, Oct. 2010.

\bibitem{IoT2}
A. Zanella, N. Bui, A. Castellani, L. Vangelista, and M. Zorzi, ``Internet of Things for smart cities,'' \emph{IEEE Internet Things J.}, vol. 1, no. 1, pp. 22-32, Feb. 2014.

\bibitem{IoT3}
P. Bellavista, G. Cardone, A. Corradi, and L. Foschini, ``Convergence of MANET and WSN in IoT urban scenarios,'' \emph{IEEE Sensors J.}, vol. 13, no. 10, pp. 3558-3567, Oct. 2013.

\bibitem{IoRT}
M. De Sanctis, E. Cianca, G. Araniti, I. Bisio, and R. Prasad, ``Satellite communications supporting Internet of remote things,'' \emph{IEEE Internet Things J.}, vol. 3, no. 1, pp. 113-123, Feb. 2016.

\bibitem{satm2m}
\emph{Satellite Machine-to-Machine (M2M) Communications---Global Strategic Business Report}, Global Ind. Anal. Inc., San Jose, CA, USA, Jul. 2016, p. 197.

\bibitem{satiot}
Z. Qu, G. Zhang, H. Cao, and J. Xie, ``LEO satellite constellation for Internet of Things,'' \emph{IEEE Access}, vol. 5, pp. 18391-18401, 2017.

\bibitem{spacex}
\emph{SpaceX Non-Geostationary Satellite System}, Federal Commun. Commissions, Washington, DC, USA, 2016.

\bibitem{oneweb}
\emph{OneWeb Non-Geostationary Satellite System}, Federal Commun. Commissions, Washington, DC, USA, 2016.
 
\bibitem{Boya}
B. Di, L. Song, Y. Li, and H. V. Poor, ``Ultra-dense LEO: integration of satellite access networks into 5G and beyond,'' \emph{IEEE Wireless Commun.}, vol. 26, no. 2, pp. 62-69, April 2019.

\bibitem{Boya2}
B. Di, H. Zhang, L. Song, Y. Li, and G. Y. Li, ``Ultra-dense LEO: integrating terrestrial-satellite networks into 5G and beyond for data offloading,'' \emph{IEEE Trans. Wireless Commun.}, vol. 18, no. 1, pp. 47-62, Jan. 2019.

\bibitem{noma}
X. Yan, K. An, T. Liang, G. Zheng, Z. Ding, S. Chatzinotas, and Y. Liu ``The application of power-domain non-orthogonal multiple access in satellite communication networks,'' \emph{IEEE Access}, vol. 7, pp. 63531-63539, 2019.

\bibitem{matching}
B. Soret, I. Leyva-Mayorga and P. Popovski, ``Inter-Plane Satellite Matching in Dense LEO Constellations,'' in \emph{Proc. 2019 IEEE Global Communications Conference}, Waikoloa, HI, USA, 2019, pp. 1-6.

\bibitem{servicetype}
3GPP, ``Service requirements for machine-type communications,''  TS 22.368 V13.1.0, Dec. 2014.

\bibitem{servicetype2}
G. Wu, S. Talwar, K. Johnsson, N. Himayat, and K. D. Johnson, ``M2M: From mobile to embedded internet,'' \emph{IEEE Commun. Mag.}, vol. 49, no. 4, pp. 36-43, April 2011.

\bibitem{ALOHA1}
G. Liva, ``Graph-based analysis and optimization of contention resolution diversity slotted ALOHA,'' \emph{IEEE Trans. Commun.}, vol. 59, no. 2, pp. 477-487, Feb. 2011.

\bibitem{ALOHA2}
E. Casini, R. De Gaudenzi, and O. Del Rio Herrero, ``Contention resolution diversity slotted ALOHA (CRDSA): An enhanced random access scheme for satellite access packet networks,'' \emph{IEEE Trans. Wireless Commun.}, vol. 6, no. 4, pp. 1408-1419, April 2007.

\bibitem{RA3}
Y. Kawamoto, H. Nishiyama, Z. M. Fadlullah, and N. Kato, ``Effective data collection via satellite-routed sensor system (SRSS) to realize global-scaled Internet of Things,'' \emph{IEEE Sensors J.}, vol. 13, no. 10, pp. 3645-3654, Oct. 2013.

\bibitem{GFRAsatellite}
R. Kassab, O. Simeone, A. Munari, and F. Clazzer, ``Space diversity-based grant-free random access for critical and non-critical IoT services.'' 2019. [Online]. Available: https://arxiv.org/abs/1909.10283v2.

\bibitem{CS}
G. Hannak, M. Mayer, A. Jung, G. Matz, and N. Goertz, ``Joint channel estimation and activity detection for multiuser communication systems,'' \emph{IEEE Inter. Conf. Commun. (ICC) Workshop}, June 2015, pp. 2086-2091.

\bibitem{CS1}
Z. Chen and W. Yu, ``Massive device activity detection by approximate message passing,'' \emph{IEEE Inter. Conf. Acoustics, Speech, Signal Processing (ICASSP)}, Mar. 2017, pp. 3514-3518.

\bibitem{CS2}
L. Liu and W. Yu, ``Massive connectivity with massive MIMO-part I: device activity detection and channel estimation,'' \emph{IEEE Trans. Signal Process.}, vol. 66, no. 11, pp. 2933-2946.

\bibitem{CS3}
L. Liu and W. Yu, ``Massive connectivity with massive MIMO-part II: achievable rate characterization,'' \emph{IEEE Trans. Signal Process.}, vol. 66, no. 11, pp. 2947-2959.

\bibitem{GQH}
Y. Zhang, Q. Guo, Z. Wang, J. Xi, and N. Wu, ``Block sparse Bayesian learning based joint user activity detection and channel estimation for grant-free NOMA systems,'' \emph{IEEE Trans. Veh. Technol.}, vol. 67, no. 10, pp. 9631-9640.

\bibitem{ZZJ}
Z. Zhang, Y. Li, C. Huang, Q. Guo, C. Yuen, and Y. L. Guan, ``DNN-aided block sparse Bayesian learning for user activity detection and channel estimation in grant-free non-orthogonal random access,'' \emph{IEEE Trans. Veh. Technol.}, vol. 68, no. 12, pp. 12000-12012.

\bibitem{ZZJAPWCS}
Z. Zhang, Y. Li, C. Huang, Q. Guo, C. Yuen, and Y. L. Guan, `` DNN-aided message passing based block sparse bayesian learning for joint user activity detection and channel estimation,'' in \emph{Proc. IEEE VTS Asia Pacific Wireless Commun. Symp.,} Singapore, 2019, pp. 1-5.

\bibitem{MF}
E. Riegler, G. E. Kirkelund, C. N. Manchon, M. Badiu, and B. H. Fleury, ``Merging belief propagation and the mean field approximation: a free energy approach,'' \emph{IEEE Trans.  Inf. Theory}, vol. 59, no. 1, pp. 588-602, Jan. 2013.

\bibitem{GMPID}
L. Liu, C. Yuen, Y. L. Guan, Y. Li, and Y. Su, ``Convergence analysis and assurance for Gaussian message passing iterative detector in massive MU-MIMO systems,'' \emph{IEEE Trans. Wireless Commun.}, vol. 15, no. 9, pp. 6487-6500.

\bibitem{GMPCW}
C. Huang, L. Liu, C. Yuen and S. Sun, ``Iterative channel estimation using LSE and sparse message passing for MmWave MIMO systems,'' \emph{IEEE Trans. Signal Process.}, vol. 67, no. 1, pp. 245-259, 1 Jan.1, 2019.

\bibitem{GMPCW2}
C. Huang, L. Liu and C. Yuen, ``Asymptotically optimal estimation algorithm for the sparse signal with arbitrary distributions,'' \emph{IEEE Trans. Veh. Technol.}, vol. 67, no. 10, pp. 10070-10075, Oct. 2018.

\bibitem{LMS}
A. Abdi, W. C. Lau, M. -S. Alouini, and M. Kaveh, ``A new simple model for land mobile satellite channels: first- and second-order statistics,'' \emph{IEEE Trans. Wireless Commun.}, vol. 2, no. 3, pp. 519-528, May 2003.

\bibitem{LMS1}
M. R. Bhatnagar, ``Making two-way satellite relaying feasible: a differential modulation based approach,'' \emph{IEEE Trans. Commun.}, vol. 63, no. 8, pp. 2836-2847, Aug. 2015.

\bibitem{LMS2}
L. Yang and M. O. Hasna, ``Performance analysis of amplify-and-forward hybrid satellite-terrestrial networks with cochannel interference,'' \emph{IEEE Trans. Commun.}, vol. 63, no. 12, pp. 5052-5061, Dec. 2015.

\bibitem{LMS3}
A. M. K., ``Two-way satellite relaying with estimated channel gains,'' \emph{IEEE Trans. Commun.}, vol. 64, no. 7, pp. 2808-2820, July 2016.

\bibitem{AccessRA}
Z. Zhang, Y. Li, L. Liu, and W. Hou, ``Fixed-symbol aided random access scheme for Machine-to-Machine communications,''  \emph{IEEE Access}, vol. 7, pp. 52913-52928, 2019.

\bibitem{EM}
J. P. Vila and P. Schniter, "Expectation-maximization Gaussian-mixture approximate message passing,'' \emph{IEEE Trans. Signal Process.}, vol. 61, no. 19, pp. 4658-4672, Oct.1, 2013.

\bibitem{modelreview}
M. S. Karaliopoulos and F. -N. Pavlidou, ``Modelling the land mobile satellite channel: a review,'' \emph{Electronics \& Communication Engineering Journal}, vol. 11, no. 5, pp. 235-248, Oct. 1999.

\bibitem{Loo}
C. Loo, ``A statistical model for a land mobile satellite link,'' \emph{IEEE Trans. Veh. Technol.}, vol. VT–34, pp. 122–127, 1985.

\bibitem{messagecombination}
H. A. Loeliger, J. Hu, S. Korl, Q. Guo, and L. Ping, ``Gaussian message passing on linear models: an update,'' \emph{Int. Symp. on Turbo codes and Related Topics}, Apr. 2006.

\bibitem{RLN}
G. E. Corazza and F. Vatalaro, ``A statistical model for land mobile
satellite channels and its application to nongeostationary orbit systems,'' \emph{IEEE Trans. Veh. Technol.}, vol. 43, pp. 738-742, Apr. 1994. 

\bibitem{DopplerElimination}
S. Amiri and M. Mehdipour, ``Accurate doppler frequency shift estimation for any satellite orbit,'' \emph{2007 3rd International Conference on Recent Advances in Space Technologies}, Istanbul, 2007, pp. 602-607.

\bibitem{viswanath}
P. Viswanath, D. N. C. Tse, and R. Laroia, ``Opportunistic beamforming using dumb antennas,''\emph{IEEE Trans. Inf. Theory}, vol. 48, no. 6, pp. 1277-1294, June 2002.

\bibitem{BGMP}
L. Liu, C. Huang, Y. Chi, C. Yuen, Y. L. Guan, and Y. Li, ``Sparse vector recovery: Bernoulli-Gaussian message passing,'' in \emph{Proc. 2017 IEEE Global Communications Conference}, Singapore, 2017, pp. 1-6.
%\bibitem{patzold}
%M. P\"atzold, U. Killat, F. Laue, and Y. Li, ``On the statistical properties of deterministic simulation models for mobile fading channels,'' \emph{IEEE Trans. Veh. Technol.}, vol. 47, no. 1, pp. 254-269, Feb. 1998.
\end{thebibliography}

\end{document}